%% file: trionpaper.tex
\documentclass[twocolumn,prb,amsmath,amssymb,superscriptaddress,floatfix]{revtex4}

\usepackage{times}
\usepackage[final]{graphicx}
\usepackage{subfigure}
\usepackage{color}
\usepackage{amsmath}
\usepackage{txfonts}
\usepackage{natbib}

\graphicspath{{img/}} 

\begin{document}

\title{Exact diagonalization study of the trionic crossover and the
trion liquid in the attractive three-component Hubbard model}
\author{Guido Klingschat}
\affiliation{Theoretical Physics, University of W\"urzburg, D-97074 Germany}
\affiliation{Institute for Theoretical Solid State Physics, RWTH Aachen University, D-52062 Aachen, Germany}
\author{Carsten Honerkamp}
\affiliation{Theoretical Physics, University of W\"urzburg, D-97074 Germany}
\affiliation{Institute for Theoretical Solid State Physics, RWTH Aachen University, D-52062 Aachen, Germany, and JARA-FIT (Fundamentals of Future Information Technology)}

\date{\today}

\begin{abstract}
We investigate the trion formation and the effective trionic properties in the attractive Hubbard model with three fermionic colors using exact diagonalization. The crossover to the trionic regime with colorless compound fermions upon increasing strength of the onsite attraction parameter $U$ features smoothly evolving ground state properties and exhibits clear similarities to the BCS/BEC-crossover for two colors. In the excitation spectrum, there is a clear gap opening between a band of well-defined trions and excitations of broken-up trions at $U_c \sim 1.8t$. This picture remains the same away from the SU(3)-symmetric point.  The spatial pairing correlations for colored Cooper pairs are compatible with a power-law at small attractions and change to an exponential decay above the trionic crossover.  Furthermore, we show that the effective trionic liquid for $U > U_c$ can be well modeled with spinless 'heavy' fermions interacting with a strong nearest neighbor repulsion.
\end{abstract}

\maketitle

\section{Introduction}
The possibility to load several hyperfine species of ultracold atoms into optical lattices \cite{ImmanuelBloch02292008} has spurred interest in the study of multi-color lattice models where the color index is used for the internal quantum number of the atoms. In particular, fermionic models with three colors were considered \cite{PhysRevLett.92.170403,PhysRevB.70.094521,PhysRevLett.98.160405,PhysRevB.77.144520,PhysRevLett.99.130406,PhysRevA.77.013624} because three colors is the first step beyond the usual two internal degrees of freedom of simple spinful many-electron systems, and, more ambitiously, these 3-color systems might reproduce phenomena known from quantum chromodynamics where 3 colors of quarks interact and give rise to different types of composite particles. Assuming an idealized situation where the hopping of the individual colors and the interaction between them are color-independent, the core model is the SU(3)-symmetric Hubbard model with the Hamiltonian
\begin{equation}
    H= - t {\sum_{<i,j>,\alpha}} c_{i\alpha}^\dagger c_{j\alpha} - \frac{U}{2} \sum_{i,\alpha\ne\beta} c^\dagger_{i\alpha} c_{i\alpha} c^\dagger_{i\beta} c_{i\beta} \; ,
    \label{eq:hamilton}
\end{equation}
with fermionic annihilation (creation) operators $c_i^{(\dagger ) }$. The color indexes $\alpha$, $\beta$ range from 1 to 3. $t$ denotes the hopping amplitude between nearest neighbored sites $i$ and $j$ on the lattice. Note that we have chosen to use the parameter values $U>0$ for the attractive case.

In the ultracold atom framework, the strength and sign of the onsite interaction $U$ between the atoms can be taken as variable. The case of attractive interactions $-U<0$ seems to offer an appealing trajectory as the strength of the attraction is increased. At weak coupling, the two-particle interaction flows to strong coupling at low temperatures, leading to the formation of Cooper pairs \cite{PhysRevLett.92.170403,PhysRevB.70.094521,PhysRevLett.99.130406}. As the bare attraction and the kinetic energy are SU(3) symmetric, there is a degeneracy between the energies of all possible two-color pairing states. In mean field theory for the infinite system where symmetry-breaking is possible, this results in a 5-dimensional sphere of degenerate BCS ground states, with a vector-like order parameter composed out of three complex numbers $\Delta_{\alpha \beta}= -\Delta_{\beta \alpha}$, where only the magnitude $\Delta_0 = \sum_{\alpha \beta} |\Delta_{\alpha
\beta}|^2/2$ is fixed. This state has some rough resemblance with the color-superfluid composed out of quark-quark-pairs in high-density QCD \cite{Hertzberg:2008wr}. Upon increasing the attraction strength, the formation of onsite-molecules of three particles with different colors, so called trions, becomes more favorable \cite{PhysRevLett.98.160405,PhysRevA.77.013624}. Indeed, using a variational ansatz \cite{PhysRevLett.98.160405}, a {\em trionic transition} from the color-superfluid into a trionic phase with colorless, heavy trions was found. The trions are the analogues of hadrons in QCD, and the famous QCD transition is supposed to occur between a colorful state at high quark densities and confined hadronic matter at lower densities. Interestingly, the QCD phase transition is usually described as first order transition, while the trionic transition in the attractive Hubbard model was first found to be continuous
\cite{PhysRevLett.98.160405}. Very recently, a refined variational study for finite temperatures indicated a first-order transition for the trionic case as well\cite{PhysRevA.80.041602}.

With the gross features of the phase diagram as function of $U$ now being clear, many interesting questions arise. Naturally, one would like to check the results of the variational approach \cite{PhysRevLett.98.160405} with an independent method and learn more about the trionic transition. Further the trionic state itself is an interesting interacting heavy-fermion system and could undergo additional phase transitions. However the quasiparticle properties and interactions of the trionic state are not easily found. One way to proceed, e.g. for finding the dispersion of the trions, is a systematic expansion in $t/U$ around the 'atomic' limit. Alternatively, the trionic bandwidth and effective interactions can be addressed using exact diagonalization (ED) of small systems which in addition provides a wealth of other information, e.g. on the excitation spectrum and correlation functions.
Here we employ full ED for the calculation of spectral functions and the Lanczos ED method to study the trion formation in the energetically lowest states. Using this information, we derive an effective low energy trion Hamiltonian. This Hamiltonian in turn can be analyzed with respect to subsequent trion phase transitions.

For the one-dimensional case, additional instabilities of three-component systems  like density waves or a Fulde-Ferrell-Larkin-Ovchinnikov (FFLO) phase have already been discussed, e.g., using density matrix renormalization group \cite{PhysRevA.80.013616,PhysRevA.77.013624,luscher-2009}. Moreover breached pairing was found for Fermi gases in a trap \cite{PhysRevA.79.051603}. There also were reports about the successful creation of a degenerate Fermi gas consisting of three different hyperfine states of $^{6}$Li (Ref. \onlinecite{PhysRevLett.101.203202}) as well as mixtures of $^{6}$Li with $^{40}$K (Ref. \onlinecite{PhysRevLett.100.053201}) or even $^{173}$Yb (Ref. \onlinecite{PhysRevLett.98.030401}). Further there are theoretical estimates on the stability of such mixtures themselves \cite{PhysRevA.77.033627} and against particle loss due to three-body recombination processes
\cite{PhysRevA.79.053633,PhysRevLett.102.165302} .

\section{Low energy spectrum and trionic regime}
\begin{figure}[htb]
    \resizebox{\columnwidth}{!}{
    \input{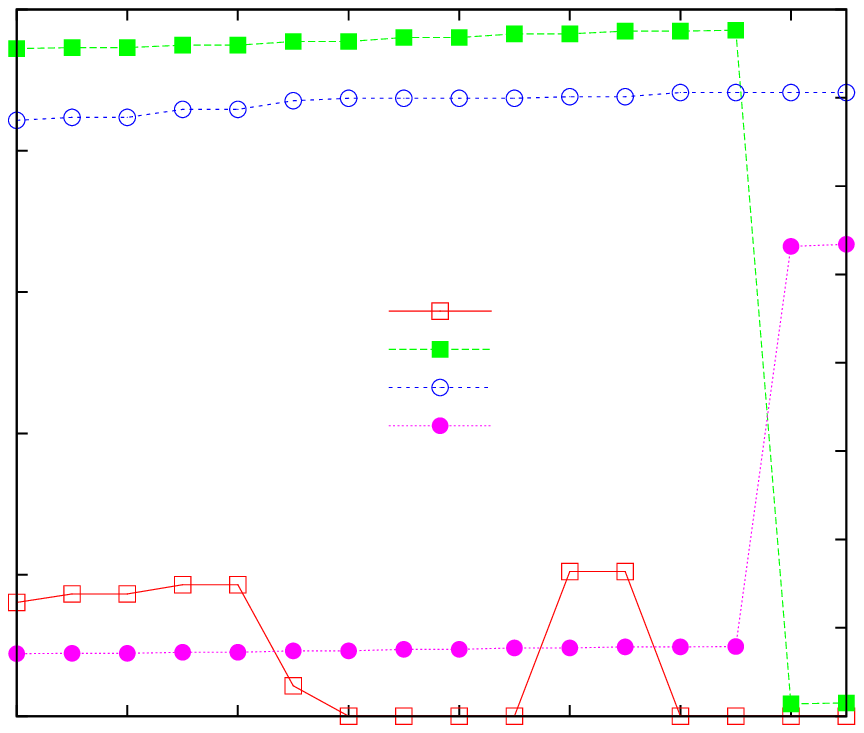}
    }
    \caption{(color online). Data for a 14-site-chain, 1 fermion/color. Right vertical scale: Empty circles: Energies of the lowest states from ED, $U=t$. Full circles: same for $U=6t$. Left vertical scale: Trionic weight $w_t$ for the same states, empty squares $U=t$, filled squares $U=6t$.}
    \label{fig:TrionWeight}
\end{figure}

Let us start with the many-particle low energy states of a $SU(3)$ symmetric system as found by ED using a Lanczos scheme. In order to get a concrete picture, we compute the spectrum of a small chain of $N$ sites with periodic boundary conditions and $n_{\alpha}=1$ fermion per color. In Fig. \ref{fig:TrionWeight} we show the energies of the energetically lowest $N+2$ states for weak and for stronger attraction.
The number of states shown is chosen $N+2$, as the interesting trion physics happens in the lowest $N$ states, where the $(N+1)$st state marks the beginning of the less trionic regime. For larger interactions, the $N$ lower states received their main weight from the basis states with the trion on one of the $N$ sites. We find the same sequence of energies for $N-1$ trions on $N$ sites, now the trion-hole has $N$ possibilities to reside on.
In addition, we plot the trionic weight $w_t$, defined as the fraction of an eigenstate that is due to pure trion states, i.e. those $N$ states with the maximum of three fermions on a single site. We observe that, for $U=t$, $w_t$ in low energy states is relatively small, and there are strong fluctuations in $w_t$ for the lowest $N+2$ states. Hence the trionic weight seems to be a less important property of the low energy states. This changes for larger attraction: for $U=6t$, there are $N$ low energy states which are separated in energy from the higher states by a gap $\Delta_t$ which grows roughly linearly with $U$ minus an offset. For these lowest states, $w_t$ is near unity and much higher than for the following states. The excited states have a lower trionic content and have hence a larger amplitude of basis states with broken-up trions. Clearly, this data shows the formation of a trionic band, which, for larger attraction, is separated from excited
states with split-up trions by an energy gap of the order $|U|$. One may also observe from the data at large $U$ that the lowest many-particle states have by no means the largest trionic weight. Rather the highest states in the trionic band are the most trionic. Clearly, this is due to a compromise between kinetic and interaction energy, i.e. the trions need to split up and sacrifice some binding energy in order to lower their kinetic energy.

\begin{figure}[htb]
    \resizebox{\columnwidth}{!}{
    \input{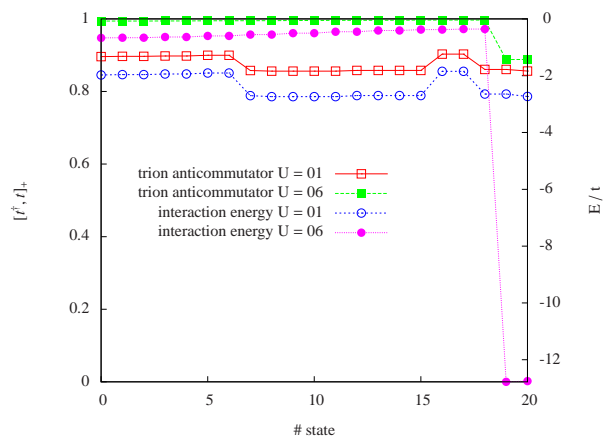}
    }
    \caption{(color online). Data for a 19-site-chain, 1 fermion/color. Right vertical scale: Empty circles: interaction energies of the lowest states from ED, $U=t$. Full circles: same for $U=6t$. Left vertical scale: Trionic anticommutator for these states, empty squares $U=t$, filled squares $U=6t$.}
    \label{fig:trionCommut-EnPot}
\end{figure}

With the definition of the trionic annihilation (creation) operator $t_{i}^{(\dagger)}$, via $t_{i}=c_{i1}c_{i2}c_{i3}$, we take a look at the spatially averaged trionic anticommutator $\left[t^{\dagger},t\right]_{+} = 1/N \sum_{i} \left( t_{i}^{\dagger} t_{i} + t_{i} t_{i}^{\dagger} \right)$,
which is a measurement for the well-definedness of the trions as compound {\em fermions}, and the binding energy of the lowest lying $N+2$ states. Now the formation of the trionic regime becomes even clearer. For weak attraction $U=t$ the lowest states in energy have nearly zero interaction energy,
that is, the expectation value of the interacting part of the Hamiltonian,
as they basically consist of non-trionic states. We furthermore see in Fig. \ref{fig:trionCommut-EnPot} that for $U=t$ the anticommutator does not change much over the $N+2$ states. It is still close to the non-interacting result $n^3+(1-n)^3 \approx 0.85$ for the density $n=1/19$ per color.
In contrast, we find jumps in these values between the $N$th and the $N+1$st state at large attraction $U=6t$. Where the binding energy of nearly $3U$ identifies the lowest $N$ states to be trionic, the drop to roughly $1U$ for the following states indicates that these are composed out of a pairs separated spatially from one single fermion. This suggestion is supported by the values of the trionic anticommutator at $U=6t$, also plotted in Fig. \ref{fig:trionCommut-EnPot}. The data clearly shows a drop between states $N$ and $N+1$.

A closer look at the anticommutator for $U=6t$ within the lowest $N$ states reflects the behavior of the trionic weight described above. While the trion formation is very strong and the anticommutator is close to one for all states in the lowest band, it is not the groundstate which has the largest anticommutator. Instead, from the groundstate to the highest trionic state the value of the anticommutator rises monotonically. This can again be explained by the fact that the states within the trionic band try to optimize their kinetic energy by breaking up the trions virtually.

\begin{figure}[htb]
    \resizebox{\columnwidth}{!}{
    \input{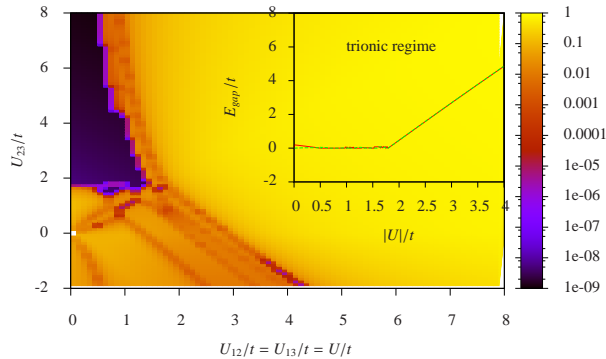}
    }
    \caption{(color online). Data for a 14-site-chain, 1 fermion/color. Location of the trion gap opening for asymmetric interactions $U_{12} = U_{13} = U > 0$ and $U_{23} \neq U$. Inset: Trionic gap vs $U/t$ for the symmetric case $U_{12}=U_{13}=U_{23}=U$.}
    \label{fig:PhaseDiag}
\end{figure}

The energy gap between the states with highest trionic weight and the rest of the spectrum shall be called the {\em trionic gap}. In the inset of Fig. \ref{fig:PhaseDiag} we plot the trionic gap $\delta_t$ as a function of $U$ for two different densities. The onset of the gap, and hence the trionic band, $U_c$ is at $2U_c \sim $bandwidth ($4t$ in this case), where $2U$ is difference of the binding energy of a single trion and a 'pair + separated-fermion' configuration. We have checked that $U_c$ for one trion on $N$ sites does not vary strongly with $N$, and also for more trions it is only changed slightly.
The value for $U_c$ agrees quite well with the location of the superfluid-to-trions transitions in the variational treatment of Rapp et al., as will be discussed more somewhat below.

The trionic gap described above shows a marked onset at a critical $U_c$. It should be emphasized however that this gap is not to be confused with a gap in the single-particle spectrum or in the effective low-energy trionic spectrum. It is a property of the excitation spectrum and not of the ground state, separating the primarily trionic many-particle states from the less trionic ones.  Hence the opening of the trionic gap should be understood as parameter value where a full band of well-defined single trion excitations becomes separated from other excitations involving broken-up trions, and where an effective trion model which does not resolve the color degree of freedom becomes appropriate.

\begin{figure}[htb]
    \resizebox{\columnwidth}{!}{
    \input{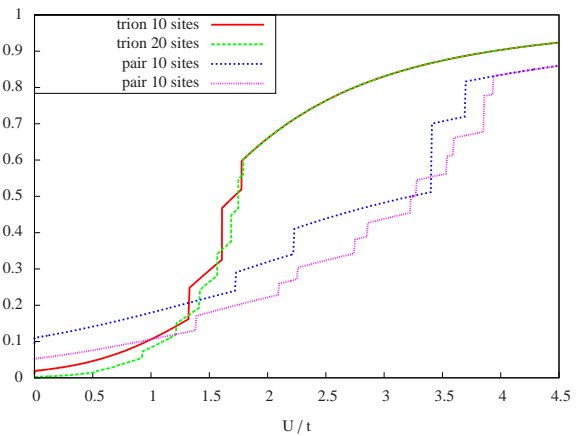}
    }
    \caption{(color online). Trionic weight and pairing weight in the ground state of 111 (short for $n_{1}=1,n_{2}=1,n_{3}=1$) and 110 systems for a 10-site-chain and a 20-site-chain.}
    \label{wtvsU}
\end{figure}

Ground state properties evolve completely smoothly as function of the interaction strength. In Fig. \ref{wtvsU} we plot the trionic weight in the ground state versus attraction strength $U$ for one trion on 10 and 20 sites. The rise of $w_t$ contains some steps which however are clearly seen to be finite-size-related. As a comparison we also plot the weight of doubly-occupied sites in the ground states in a system with only two colors and local attractions, corresponding to the BCS/BEC-crossover. As the local pairs have less binding energy, the crossover occurs at larger attraction $U$, but otherwise the behavior of the weight of compound particles in the ground state is very similar. Hence, if we do not allow for spontaneous symmetry breaking, there is a smooth connection between the weak-coupling ground state and the trionic state at large attractions. Nevertheless the opening of the trionic gap is a significant phenomenon, as it allows us to describe the low-energy theory in terms of well-separated trionic excitations only.

In Fig. \ref{fig:PhaseDiag} we also show the location of the trionic
gap opening for interactions which violate the SU(3) symmetry and where $U_{23}$ acting between colors 2 and 3 is different from the other two interaction constants.
The trion formation is a generic feature as long as all three colors attract.
A weak onsite repulsion between two colors is also tolerable, as one can see from the extension of the trionic regime to negative $U_{23}$. Hence, while the color-superfluid at weak attraction requires a symmetric situation, the trion formation is a general feature of the three-color system and could thus be realized much more easily in experiments.

We have also compared the value for the trionic gap opening for different system sizes with one or two trions and symmetric interactions. We found small quantitative changes, but there was always a clear energetic separation between primarily trionic and other states at interaction strengths above $U_c \sim$ bandwidth$/2$. The data suggests clearly that $U_c$ remains nonzero in the dilute limit.



An obvious question is whether the system of three colors of fermions with attractive interactions also contains a BEC regime, where the ground state is composed primarily out of strongly bound pairs of two different colors. In the limit of local pairs, there is a simple argument to understand that such pairs are not stable. If such a difermionic regime would exist, it should be found at $U$-values below the trionic regime, i.e. at $U\le U_c < 2dt$ in $d$ dimensions on the hypercubic lattice. Now, the binding energy of one trion is $3U$, which is $2U$ more than for a pair. So breaking up a trion costs $2U$, where the kinetic energy gain for the system is about $2dt$ for the freed fermion. Breaking up the pair as well now only costs $U$, but gains $4dt$ of kinetic energy. The latter gain is certainly larger if we are below the threshold for trions. Therefore this pair-split will be preferred instead of breaking up another trion. Accordingly there is no stable difermionic regime and by that no BEC regime with such strongly bound pairs. This can also be inferred from comparing the curves for the trionic case with three colors and the local pair case with only two colors in Fig. \ref{wtvsU}. The local pair regime occurs only deep in the parameter range of the trionic regime.

On the other hand, as discussed in previous works \cite{PhysRevA.73.053606} at weak $U$ Cooper pairing with loosely bound pairs is possible. The argument stated above basically means that the trajectory known from two-color system with increasing attraction where the pair size shrinks continuously becomes unobservable. Already at intermediate $U$, it is more favorable to form local trions, and the non-trionic many-particle states necessary for maintaining longer-ranged pairing loose their weights in the low-energy sector.

\section{Connection to the superfluid-to-trions transition and pairing correlations}
The variational analysis of Rapp et al. \cite{PhysRevLett.98.160405} showed a continuous quantum phase transition between a color superfluid at small attraction to a colorless trionic phase at larger attraction at a critical $U_{c} \sim 1.774$. This transition is driven by the competition between the Cooper pairing instability and the local formation of trions that are implemented as completely immobile or frozen in the variational ansatz of \cite{PhysRevLett.98.160405}.
In this trionic wave-function, the expectation value for any non-local pairing correlation (PC) vanishes identically. This forces the BCS amplitude to precisely zero when the frozen-trions wave-function becomes energetically favorable.
Very interestingly, in a recent paper Inaba and Sugo\cite{PhysRevA.80.041602} present a variational cluster approximation study of this transition for the density of states of the infinite-dimensional Bethe lattice. They find a first order transition between the color superfluid and the trionic state at low temperatures, hence the precise nature of this transition is at least a subtle issue that in theory depends on the approximation.

In our finite systems, there cannot be any spontaneous symmetry breaking. Hence instead of a quantum phase transition we rather have a crossover into the trionic regime. Nevertheless one can expect that the $U$-values for the formation of the well-separated trion band are a good estimate for the attraction necessary to suppress the color superfluid efficiently by strongly reducing the weight of non-trionic states in the ground state. Indeed, this value, expressed in $U$ over the bandwidth, from the exact diagonalization in one and two dimensions is very close to the variational results in the limit in infinite dimensions and to the value in the VCA study\cite{PhysRevA.80.041602}.

\begin{figure}[htb]
    \begin{center}
        \resizebox{\columnwidth}{!}{
        \input{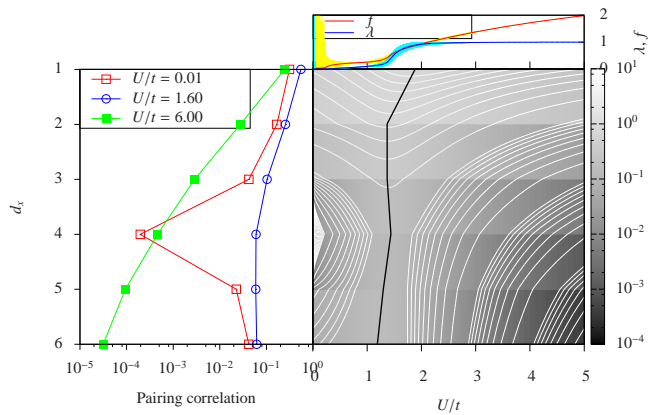}
        }
    \end{center}
    \caption{(color online). Data for a 12-site-chain, 333 filling. Left: logarithmic plot of the pairing correlations $\Pi$ vs. distance $d_x$ in the groundstate for different $U$. For very weak interaction ($U=0.01$, open squares) the data shows a dip a $k_Fd_x=\pi$ as expect from the non-interacting case. One can also observe the nearly exponential decay for strong interactions ($U=6.00$, filled squares). For intermediate attraction ($U=1.60$, open circles) near the crossover to trions the pairing correlations are largest.
Right: logarithmic mapping of the pairing correlation for distances up to half the system size and attractions $0 \leq U \leq 5$ and contour lines. The solid line shows the $U$-dependence of the pairing correlations maximum. Top: fitting parameter $\lambda$ (with error bars) discriminating the weakly-interacting regime from the regime with exponential decay of pairing. Also shown are the fit results for the exponential decay factor $f$, which have huge error bars for small interaction where $\lambda$ is near zero and the exponential decay plays only a negligible role.}
    \label{fig:1D-pairing-12333}
\end{figure}

The spatial Cooper paring correlations, for simplicity in a one-dimensional chain,
\[ \Pi (x-x') = \sum_x \langle c_\alpha (x) c_\beta (x)  c^\dagger_\beta (x')  c^\dagger_\alpha (x')\rangle_{\mathrm{GS}} \]
for $\alpha\not=\beta$ can be readily measured in the ground state found by the diagonalization.
In the groundstate of a non-interacting Fermi liquid one finds a spatial dependence that goes like
\[ \Pi_0 (\Delta x) \sim \frac{ \sin^{2}(k_{F}\Delta x) }{ (\Delta x)^{2} }\]
with the distance $\Delta x = x-x'$. This gives a node at distance $\Delta x = \pi / k_{F}$ with the Fermi vector $k_{F}$. For a system with three fermions of each species on a 12-site-chain the node is expected at $\Delta x = 4$.
In the strongly coupled regime one expects an exponential decay of the pairing correlations, as doubly occupied sites necessary to maintain the pairing are punished by a finite energy $\sim2U$, and hence the weight of these states in the ground state should be suppressed exponentially.
The left part of Fig. \ref{fig:1D-pairing-12333} shows the pairing correlations for different $U$ as function of the distance.  For very small interactions, the node of the free solution is clearly visible.
In the trionic regime we find an exponential decay realized up to $\Delta x = k_{F}/\pi = 4$. For larger distances the decay is slower due to finite size effects coming in from the other side. The data also shows that for intermediate attraction neither of the two descriptions is good and one gets a crossover behavior. In fact one would expect power-law behavior, but we refrain from determining an exponent in our small systems. The main point is the onset of an exponential decay above a critical attraction strength.

The right part of Fig. \ref{fig:1D-pairing-12333} displays the logarithm of the pairing correlations on a distance-vs-interaction grid together with the respective contour lines. The solid ''vertical'' line sketches the $U$-dependent maximum of the pairing correlations. The data shows clearly that a small onsite attraction first enhances the Cooper pairing. Above a threshold value below the onset of the trion formation, the Cooper pairing is suppressed. This behavior of the pairing correlations vs $U$ for our small 1D system follows closely the variational result for the infinite and high-dimensional system of Rapp et al. \cite{PhysRevLett.98.160405}.
Fitting the data with a linear combination of the two models
\begin{align}
    \begin{split}
        \Pi (x) &\sim \lambda \cdot \left( e^{-f x} + e^{-f (N-x)} \right)           \\
        &+ (1 - \lambda) \cdot \left( \Pi_{0}(x) + \Pi_{0}(N-x) \right)
    \end{split}
    \label{eq:PC-lincomb}
\end{align}
where $f$ is a free fit parameter and $0 \leq \lambda \leq 1$ interpolates between free fermion behavior ($\lambda=0$) and exponential decay ($\lambda=1$). Note that the rise of the weight $\lambda$ of the exponentially decaying component is rather steep. It seems well possible that in a larger system the rise will become a step. This would be consistent with a first order transition. In any case the data indicates that the pairing correlations undergo a rapid change as function of the attraction strength,  making it very plausible that the precise nature of the color-symmetry breaking transition in the infinite system is a subtle issue.

We think that the similarity of our data for the small chain to the variational results for the thermodynamic limit in Refs. \onlinecite{PhysRevLett.98.160405,PhysRevA.80.041602} is due to the local nature of the trion formation which is to a large degree independent of the dimensionality or lattice topology. Nevertheless one might wonder if some kind of longer-ranged pairing correlations could persist even in the trionic regime of the infinite system, as indicated by our fit to the pairing correlations that shows a remnant power-law above the transition.  Hence it is interesting to investigate  the superfluid-to-trions transition in a framework that goes beyond the treatment of Ref. \onlinecite{PhysRevLett.98.160405} by allowing a certain admixture of non-trionic states into the ground state that could support Cooper pairing even in the trionic regime. The VCA study of Ref. \onlinecite{PhysRevA.80.041602} is a first step, but it still works in infinite dimensions where the kinetic energy of the trions is zero due to the scaling of the single-fermion hopping. This could underestimate the admixture of non-trionic states to the low energy sector.

\section{Effective trion Hamiltonian}
As the trions are relatively well-defined anti-commuting, i.e. fermionic particles for sufficiently large $U$, we can attempt to describe the low-energy physics with an effective trion Hamiltonian. In principle, the effective trion hopping and also the effective interactions can be found analytically by a $t/U$ expansion \cite{toke}. The main contributions were already given qualitatively in Ref. \onlinecite{PhysRevB.77.144520}.
Alternatively, with our ED approach, the values for hopping parameters can be simply read off from the eigenvalues of a small system of $N$ sites with one fermion per color, i.e. one trion. The $N$ lowest states for $U_c$ in the case of a single trion in the system can be associated with $N$ momentum eigenstates and nicely fit on a $-2 t_{\mathrm{eff}} \cos k_x$-curve, i.e. nearest neighbor hopping dominates at least well in the trionic regime. From this we can read off the bandwidth $D_t=4t_{\mathrm{eff}}$ of the effective trions band. We find $D_t= 1.5 t^3/U^2$ for sufficiently large $U$. The third power of $t$ occurs due to the three single fermion hoppings necessary to move a trion as a whole by one site. As our data are obtained on small systems one might worry if a finite-size gap $\sim 1/$length spoils the detection of the small kinetic energy scale for the trion. Note however that the relevant finite-size gaps, measured by the difference between neighboring energy eigenvalues within the trionic band, also scale down with $t^3/U^2$, i.e. always remain smaller than the effective bandwidth.

\begin{figure*}[htb]
    \begin{center}
        \subfigure{
        \label{fig:trioncorrelation-1d+2d}
        \resizebox{\columnwidth}{!}{
        \input{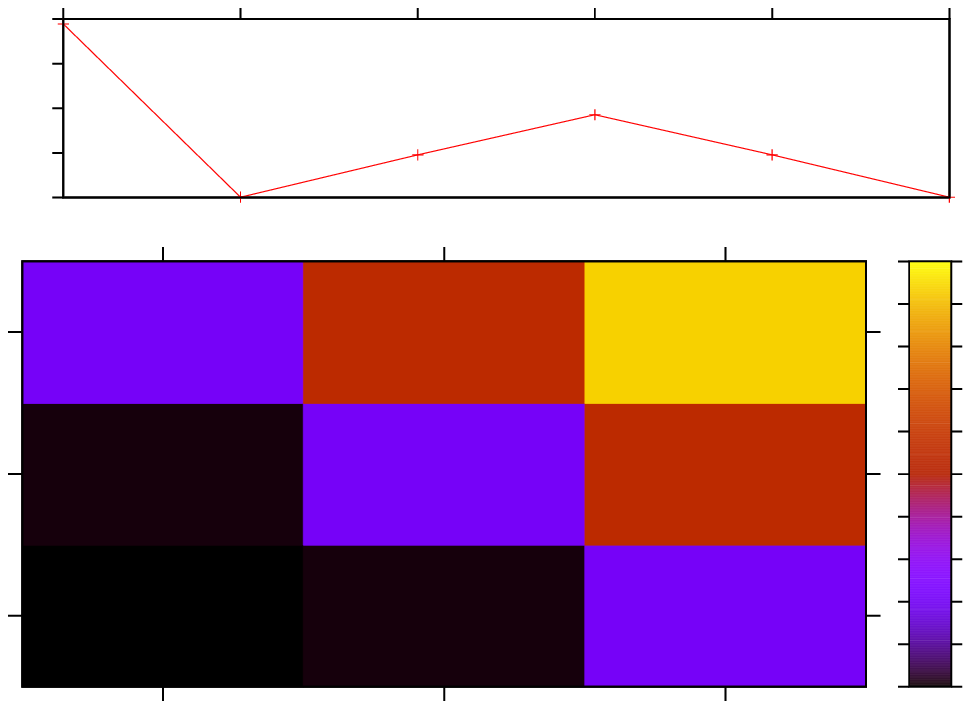}
        }
        }
        \subfigure{
        \label{fig:diff-trion-fermion-correlation}
        \resizebox{\columnwidth}{!}{
        \input{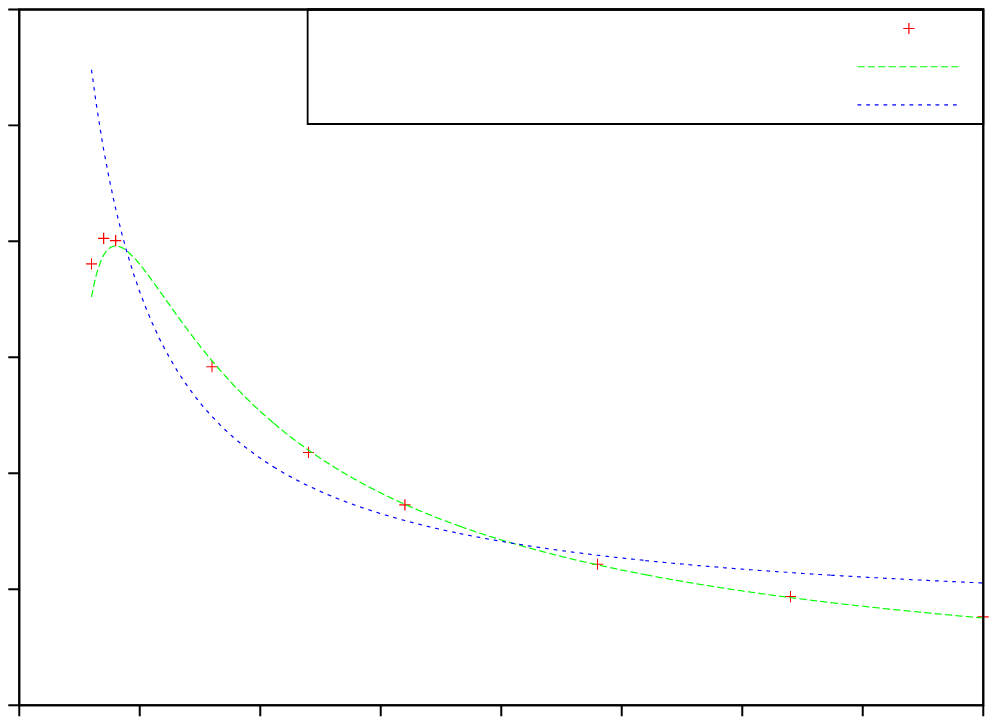}
        }
        }
    \end{center}
    \caption{
    (color online).
    Left Top: Data for a 6-site-chain, 2 fermions/color, $U=8t$. The spatial trionic correlation function shows a strong next neighbor repulsion. The value of $\langle n^{t}(0)n^{t}(0) \rangle < 1$ reflects the circumstance that the trions are composite particles that are broken up to some degree by quantum fluctuations.
    Left Bottom: Data for a 4$\times$4 square cluster, 2 fermions/color, $U=8t$. Trionic correlation function shows strong next-neighbor repulsion, similar to expectations for spinless fermions.
    Right: Enhanced expectation value for trions on (2,2) relative to expectation for spinless fermions and $U$-dependence of enhancement. Best fit uses a $U^{-1}$-term plus an additional $U^{-2}$-term with negative prefactor and has excellent agreement with the data.
    }
    \label{fig:trioncorrelation-U08-4x4-222-contour}
\end{figure*}

The effects of the interaction between the trions can be readily observed in the spatial trion correlation functions $\langle n^{t}(x) n^{t}(x') \rangle$, where $n^{t}(i) = t^{\dagger}_{i} t_{i}$ tells if there is a trion on site $i$ or not.
As visible in the top part of Fig. \ref{fig:trioncorrelation-U08-4x4-222-contour} the main term is a strong nearest neighbor repulsion. The physical origin of this is the suppression of a kinetic energy gain of the order $t^2/U$ due to single fermion site fluctuations if two trions sit next to each other. This interaction $V_1 \sim t^2/U$ is already comparable to the effective bandwidth at $U_c$ which grows with respect to the bandwidth deeper into the trionic regime. Hence the trionic liquid is an intrinsically strongly interacting system of heavy composite fermions.

\begin{figure}[htb]
    \subfigure[$U/t=4$]{
    \label{1d-trion-spec-U04}
    \resizebox{\columnwidth}{!}{
    \input{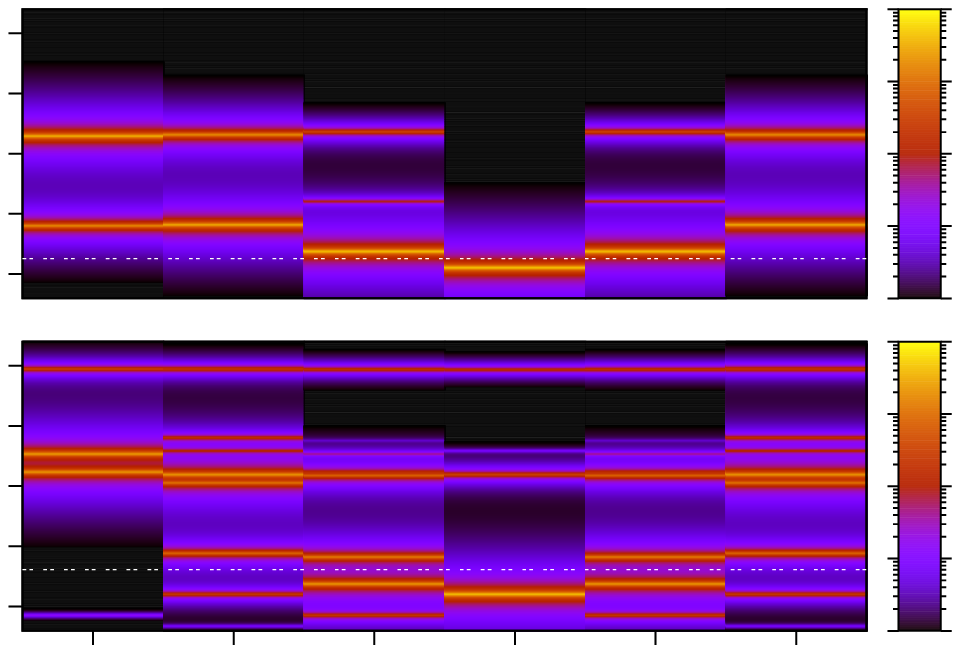}
    }
    }
    \subfigure[$U/t=8$]{
    \label{1d-trion-spec-U08}
    \resizebox{\columnwidth}{!}{
    \input{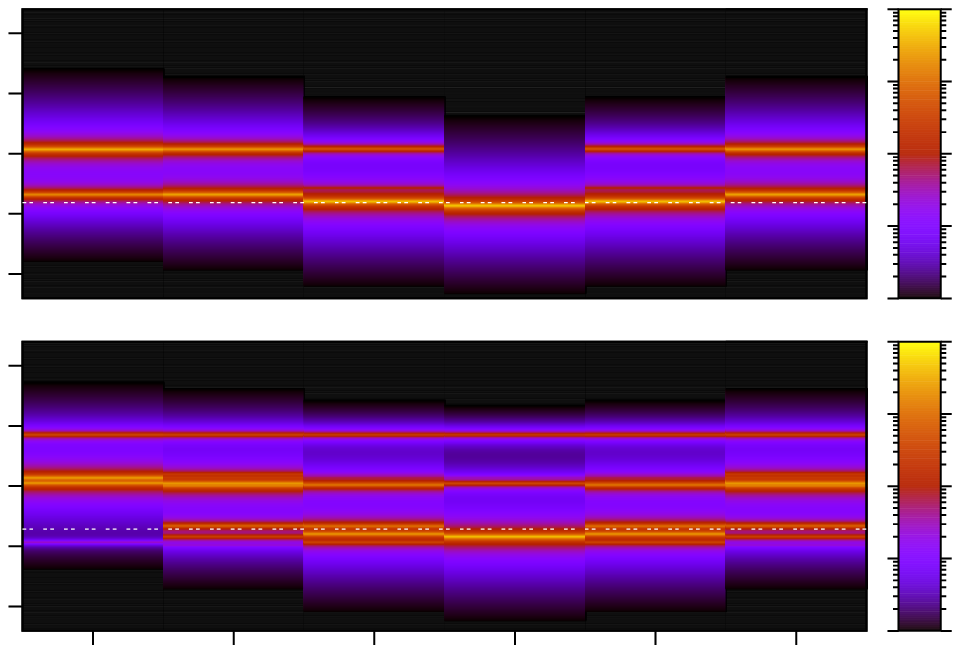}
    }
    }
    \caption{(color online). Data for a 6-site-chain, 1 (2) fermion/color for respective upper graph of a (b). The dashed line marks the chemical potential $\mu$. The trionic spectral function shows a cosine band of trion states. This band shows in the lower plots in (a) and (b) a gap opening above the Fermi level at $|k|=\pi/2$ and some backfolding for the case of two trions. This gap (in the lower plot of (a) between $\omega \approx -12.5t$ and $\omega \approx -12t$) scales with $t^2/|U|$ and is larger than typical finite size effects of the system measured by energy differences in the band between neighbored $k$-values. The width of the trion band with the gap subtracted scales as $t^3/U^2$. There is an additional split-up of the $k=\pm \pi/3$ excitations that seems to scale in the same way as the bandwidth. }
    \label{1d-trion-spec}
\end{figure}

What are the consequences of these interactions?
First, near half band filling on a bipartite lattice, $V_1$ will favor a density wave (DW) ground state of the trions with alternating densities from site to site. This state is adiabatically connected to the density wave state of the colored fermions at weaker coupling \cite{PhysRevLett.92.170403}. In one dimension, this CDW for trions has been found numerically by Molina et al. \cite{PhysRevA.80.013616} and Azaria et al. \cite{PhysRevA.80.041604}. Moving the density away from half filling, the DW order will melt, possibly with an intermediate regime of phase separation. Yet, at one-third filling, it can be seen that $V_1$ still strongly influences the trion system even in absence of the DW. The trionic Green's function is
defined in real space by
\begin{equation*}
    G_{ij}(\omega) = \sum\limits_{n} \frac{\langle 0 | t_j^{\dagger} | n \rangle  \langle n | t_i | 0 \rangle} {\omega - (E_{0}-E_{n}) + i\,\eta^{+}}
    + \sum\limits_{m} \frac{\langle 0 | t_i | m \rangle  \langle m | t_j^{\dagger} | 0 \rangle} {\omega-(E_{m}-E_{0}) + i\,\eta^{+}}
    \label{eqn:spektraldarstellung-greenfunktion-T=0}
\end{equation*}
with sums over intermediate states $n, m$ and numerical broadening $\eta$. The first term corresponds to the emission spectrum, i.e. transitions from the ground state into states with one trion less, while the second part is the inverse emission spectrum.  The wavevector-resolved spectral function is the imaginary part of the Fourier-transform of $G_{ij} (\omega )$  to wavevector $k$. This is shown in Fig. \ref{1d-trion-spec}.
The lower plots in (a) and (b) are for two trions in the system. The $-2 t_{\mathrm{eff}} \cos k$- trionic band exhibits a marked break-up $\propto t^{2}/|U|$  between portions concentrated at $|k| >  \pi/2$ and portions connected to $k=0$ above the trionic Fermi level. Here the Fermi level is determined by the energy where the emission part ends and where the inverse emission part starts, it intersects the band at $k = \pm \pi/3$. The half-filled $k= \pm \pi/3$-states are slightly split up as well. In the upper plots of Fig. \ref{1d-trion-spec} (a) and (b) we show the spectral function for just one trion in the system. Now one can see a continuous trionic band, but the inverse photoemission part contains another quasiband shifted upward by an energy $\propto t^2/U$, which is a signature of the interaction of the added trion with the trion from the initial state.

In fact, the many-particle spectrum at low energies with the break-up of the trion band can be nicely reproduced with spinless fermions with effective hopping $t_t$ and nearest neighbor interaction $V_1$. This comparison allows us to get the value for the effective $V_1$ of the trions. We therefore first simulated such a spinless fermion system, represented by the Hamiltonian
\begin{equation} H= - \tilde t {\sum_{\langle i,j\rangle}} \left( c_{i}^\dagger
    c_{j}+ c_{j}^\dagger
    c_{i} \right) + \tilde V \sum_{\langle i,j \rangle} c^\dagger_{i} c_{i}
    c^\dagger_{j}
    c_{j} \; , \label{eq:hamilton-spinless}
\end{equation}
with one spinless fermion on 16 sites and compared the width of the spectrum $\Delta E$ with the $U$-dependent bandwidth of the according trionic system.
\begin{equation}
        t_{\mathrm{eff}}(U) = \frac{\Delta E}{4} = y \frac{t^{3}}{U^{2}}
    \label{eq:hopping-effective}
\end{equation}
The proportionality factor results to $y \leq 1.50$ where the equality holds in the limit of infinitely strong attraction $U$.
With this value for the effective trionic hopping we compared the gap within the trionic band that occurs due to the strong trionic interactions for more than one trion in the system with the gap in the spectrum of the spinless fermions which becomes $\tilde \Delta \approx \tilde V + 2 \tilde t \tilde{n} $ for large enough $\tilde V$. The second term is a  kinetic energy loss due to the partial Pauli-blocking of neighbored sites depending on the density of the spinless fermions, $\tilde n$.
For a system of two spinless particles on twelve sites this gap opens between the 54th and the 55th lowest state. The 3-color trionic system develops a gap at the same position, given by
\begin{equation}
    \Delta_{t} = 2 \frac{U}{t} t_{\mathrm{eff}} + 2 t_{\mathrm{eff}} n_{t} .
    \label{eq:repulsion-comparison}
\end{equation}
    By equating the first term  to the nearest neighbor interaction of the effective trion Hamiltonian, we get the very reasonable result $V_{\mathrm{eff}} =  2 \frac{U}{t} t_{\mathrm{eff}} = 3 t^2/U$ for large enough $U$.

\begin{figure}[htb]
    \begin{center}
        \resizebox{\columnwidth}{!}{
        \input{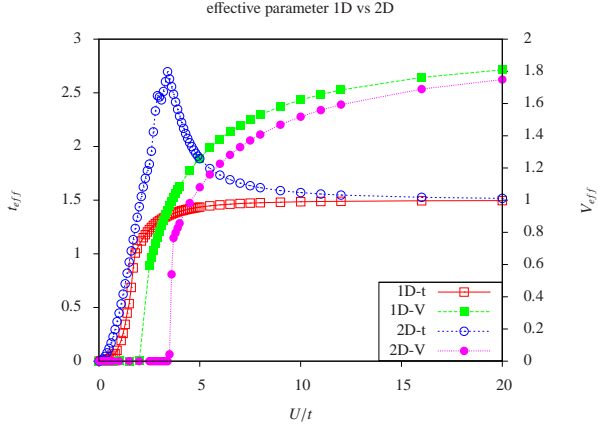}
        }
    \end{center}
    \caption{(color online). Data for a 16-site-chain vs 4$\times$4 square cluster, 111 filling for effective hopping (left vertical scale), 222 filling to find effective repulsion (right vertical scale). Left vertical scale, open symbols: the 2D hopping (circles) shows an extra term compared to the 1D hopping (squares). Right vertical scale, filled symbols: the effective repulsion in 2D (circles) indicates the larger $U_{c}$ for the trion regime than in 1D (squares).}
    \label{fig:paramsvsU}
\end{figure}

In summary, we find, including terms up to $t^3/U^2$, the following effective Hamiltonian for the trionic sector:
\begin{align}
    \begin{split}
        H_{\mathrm{\mathrm{eff}}} =
        &- \frac{3}{2} \frac{t^{3}}{U^{2}} \sum_{\langle i,j \rangle } \left( t_i^{\dagger} t_j + t_j{^\dagger} t_i \right)              \\
        &+ \frac{3}{2}\frac{t^{2}}{U} \sum_{\langle i,j \rangle } n^{t}_{i} n^{t}_{j}
    \end{split}
    \label{eq:heff}
\end{align}
From these terms we observe that there is a hierarchy of energy scales: the dominant term is the nearest neighbor interaction $V_{\mathrm{eff}} \sim t^2/U$, where $|t/U|<1$. The trion hopping is of order $ t^3/U^2$. In this Hamiltonian, higher order corrections have not been written. In Fig. \ref{fig:paramsvsU} we display the convergence of trion hopping (or bandwidth respectively) and interactions toward these values as function of $U$. Notably, on the 2D square lattice the corrections are stronger than in one dimension.

When we extend our analysis to a 2-dimensional square lattice, e.g. a square 4$\times$4 cluster, the behavior of the trions remains qualitatively the same. As the hopping is now allowed in more than one direction, the width of the trion band and the critical value of $U$ before a trionic gap opens are roughly doubled.

The effective nearest neighbor hopping for the trions versus $U$ is shown in Fig. \ref{fig:paramsvsU}. It approaches the same constant $t_{\mathrm{eff}} = 1.5 t^3/U^2$ for large $U$. One however observes that with increasing $U$ the data in 2D converges from above to this value, while it comes from below in 1D. The difference is due to an additional higher order hopping contribution $\sim t^5/U^4$ coming from two color fermions hopping directly to the neighbored site and 1 color fermion hopping around the plaquette. Longer ranged trion hoppings like the one across the diagonal also exist, but decay at least as $t^6/U^5$ and are hence less important.

Fitting the effective hopping dispersion in two dimensions with a model including nearest-neighbor ($a$) , (linear) next-nearest-neighbor ($b$) and diagonal hopping ($c$),
\begin{align}
    \begin{split}
        t_{\mathrm{eff}}(k_x,k_y)/t   &= const                                    \\
        &+ a \cdot \left( \cos(k_{x}) + \cos(k_{y}) \right)          \\
        &+ b \cdot \left( \cos(2 k_{x}) + \cos(2 k_{y}) \right)      \\
        &+ c \cdot \left( \cos(k_{x}+k_{y}) + \cos(k_{x}-k_{y}) \right)
    \end{split}
    \label{eq:teff}
\end{align}
we get the following results for the $u = U/t$-dependence of the parameters
\begin{align}
    \begin{split}
        const &= - 3 u - 3 u^{-1}                                    \\
        a &= - 3 u^{-2} - 12 u^{-4}   \\
        b &= - 6.7 u^{-5} - 12 \cdot 6.7 u^{-7}    \\
        c &= 2 \cdot b \, .
    \end{split}
\end{align}
Here the prefactors of the second leading order terms in $a$ and $b$ are not the result of best fit procedures, but lead to reasonably small deviation from the numerical data using integer numbers and simple fractions. The prefactor 12 in front of the $U^{-7}$-term in $b$ and especially $c=2b$ are exact within numerical precision. This factor of 2 for the diagonal hopping with respect to the linear next-to-nearest neighbor hopping can be understood easily as there are two ways around the corner to the diagonal site.

As one would have expected, the spatial trion correlations on the 2D square lattice exhibit again signatures of the effective strong nearest neighbor repulsion. This can be clearly seen on the 4$\times$4 plaquette with periodic boundary conditions shown in Fig. \ref{fig:trioncorrelation-U08-4x4-222-contour}. The nearest neighbor sites (1,0) and (0,1) have basically zero probability for both being occupied if there is a trion on site (0,0). Considering second nearest neighbors, the diagonal neighbors (1,1) and the linear second nearest neighbors (2,0) and (0,2) are equally populated. For larger clusters we would expect to find enhanced probability for trions on the linear neighbors instead of the diagonal ones. This is based on the following idea. As a fluctuating trion on site (1,1) has two neighbors (1,0) and (0,1) that are affected by Pauli blocking mediated by the fluctuations of the trion at (0,0), while on sites (2,0) or (0,2) only fluctuations on
one neighbored site are partially Pauli blocked. This can be confirmed for spinless fermions, but for the trions the largest system we can handle here is 4$\times$4, where e.g. (2,0) is only one hop away from (-1,0) and is hence also confronted by Pauli blocking on two sides, and there is no difference in the density correlations on these sites.

While at first sight, the spatial trion correlations look very much like the correlations of spinless fermions with appropriately chosen nearest-neighbor interactions, a detailed comparison shows that for two trions on the small cluster, the second particle is pushed away more strongly than for two spinless fermions. The bottom of Fig. \ref{fig:trioncorrelation-U08-4x4-222-contour} quantifies this difference. The expectation value for finding the second fermion on site (2,2) in the spinless fermion model is subtracted from the normalized value for finding the second trion on site (2,2). This difference decreases roughly like $U^{-1}$, supporting a $t$-$V$ model as effective model for large $U$, but showing that for intermediate $U$ the three-color model is still more complex. The best fit of this discrepancy needs an extra $U^{-2}$ term and is $\propto (U^{-1} - 4 U^{-2})$. Note that longer-ranged effective hoppings not included in the spinless-fermion model cannot be the reason for this difference, as they only start with $U^{-5}$.

\section{Conclusions} With an exact diagonalization study we have identified the trionic regime of attractive three-color fermion systems on simple one- and two-dimensional lattices. This regime develops out of the weak-coupling state with deconfined colored fermions through a smooth crossover. The amplitude of non-trionic basis states in the ground state becomes smaller with increasing attraction, but never reaches zero at any finite $U$. While the ground state properties evolve smoothly, in the excitation spectrum one finds a gap opening at a critical $U \le $ half the bandwidth, where a band on trionic single-particle excitations gets separated from less trionic excitations. This signals the onset of a regime with well-defined trions throughout the Brillouin zone. We have also shown that the Cooper pairing correlations go through a maximum slightly below the trionic gap opening when $U$ is increased but get markedly suppressed in the trionic regime. This
finite-size picture is consistent with the previous variational study of the color-superfluid-to-trion transition by Rapp et al. \cite{PhysRevLett.98.160405}. On the other hand, our data shows that the description of the trionic state should be extended to capture the admixture of broken-up-trion states to the ground state. These states are necessary for the dynamics of the trions and might support remnant Cooper pairing correlations even in the trionic regime.
In this regime, the trions can be described by an effective model of spinless heavy fermions with bandwidth $\sim t^3/U^2$ and with a strong nearest neighbor repulsion $\sim t^2/U$. We have determined the prefactors for these effective parameters. In this sense, the trion liquid is an intrinsically strongly interacting limit wherever the trions are well-defined. In the limit of large $U$, the trions can be described by a lattice gas where the particles avoid being nearest neighbors and where the kinetic energy is suppressed by a factor of order $t/U$.

The heavy trion liquid phase itself should be an interesting laboratory for many-particle physics. As already pointed out in the literature, the trions order in a density wave pattern on bipartite lattices when there is half a trion per site. Furthermore, as interacting (quasi-)fermionic particles, the trions should undergo superfluid pairing transitions in the odd-parity channel at low temperatures, reminiscent of neutron or proton superfluid in astrophysics. This will be an interesting topic for further research.

We thank M. Aichhorn, W. Hofstetter, A. L\"auchli, J. Ortloff, A. Rapp, C. Toke, and G. Zarand for useful discussion. This work was supported by the German Research foundation (DFG).

\bibliography{bib/paper}

\end{thebibliography}

\end{document}

%% file: img/trionWeight-EiVal-14111.tex
\begingroup
  \makeatletter
  \providecommand\color[2][]{%
    \GenericError{(gnuplot) \space\space\space\@spaces}{%
      Package color not loaded in conjunction with
      terminal option `colourtext'%
    }{See the gnuplot documentation for explanation.%
    }{Either use 'blacktext' in gnuplot or load the package
      color.sty in LaTeX.}%
    \renewcommand\color[2][]{}%
  }%
  \providecommand\includegraphics[2][]{%
    \GenericError{(gnuplot) \space\space\space\@spaces}{%
      Package graphicx or graphics not loaded%
    }{See the gnuplot documentation for explanation.%
    }{The gnuplot epslatex terminal needs graphicx.sty or graphics.sty.}%
    \renewcommand\includegraphics[2][]{}%
  }%
  \providecommand\rotatebox[2]{#2}%
  \@ifundefined{ifGPcolor}{%
    \newif\ifGPcolor
    \GPcolortrue
  }{}%
  \@ifundefined{ifGPblacktext}{%
    \newif\ifGPblacktext
    \GPblacktexttrue
  }{}%
  \let\gplgaddtomacro\g@addto@macro
  \gdef\gplbacktext{}%
  \gdef\gplfronttext{}%
  \makeatother
  \ifGPblacktext
    \def\colorrgb#1{}%
    \def\colorgray#1{}%
  \else
    \ifGPcolor
      \def\colorrgb#1{\color[rgb]{#1}}%
      \def\colorgray#1{\color[gray]{#1}}%
      \expandafter\def\csname LTw\endcsname{\color{white}}%
      \expandafter\def\csname LTb\endcsname{\color{black}}%
      \expandafter\def\csname LTa\endcsname{\color{black}}%
      \expandafter\def\csname LT0\endcsname{\color[rgb]{1,0,0}}%
      \expandafter\def\csname LT1\endcsname{\color[rgb]{0,1,0}}%
      \expandafter\def\csname LT2\endcsname{\color[rgb]{0,0,1}}%
      \expandafter\def\csname LT3\endcsname{\color[rgb]{1,0,1}}%
      \expandafter\def\csname LT4\endcsname{\color[rgb]{0,1,1}}%
      \expandafter\def\csname LT5\endcsname{\color[rgb]{1,1,0}}%
      \expandafter\def\csname LT6\endcsname{\color[rgb]{0,0,0}}%
      \expandafter\def\csname LT7\endcsname{\color[rgb]{1,0.3,0}}%
      \expandafter\def\csname LT8\endcsname{\color[rgb]{0.5,0.5,0.5}}%
    \else
      \def\colorrgb#1{\color{black}}%
      \def\colorgray#1{\color[gray]{#1}}%
      \expandafter\def\csname LTw\endcsname{\color{white}}%
      \expandafter\def\csname LTb\endcsname{\color{black}}%
      \expandafter\def\csname LTa\endcsname{\color{black}}%
      \expandafter\def\csname LT0\endcsname{\color{black}}%
      \expandafter\def\csname LT1\endcsname{\color{black}}%
      \expandafter\def\csname LT2\endcsname{\color{black}}%
      \expandafter\def\csname LT3\endcsname{\color{black}}%
      \expandafter\def\csname LT4\endcsname{\color{black}}%
      \expandafter\def\csname LT5\endcsname{\color{black}}%
      \expandafter\def\csname LT6\endcsname{\color{black}}%
      \expandafter\def\csname LT7\endcsname{\color{black}}%
      \expandafter\def\csname LT8\endcsname{\color{black}}%
    \fi
  \fi
  \setlength{\unitlength}{0.0500bp}%
  \begin{picture}(7200.00,5040.00)%
    \gplgaddtomacro\gplbacktext{%
      \csname LTb\endcsname%
      \put(1078,704){\makebox(0,0)[r]{\strut{} 0}}%
      \put(1078,1518){\makebox(0,0)[r]{\strut{} 0.2}}%
      \put(1078,2332){\makebox(0,0)[r]{\strut{} 0.4}}%
      \put(1078,3147){\makebox(0,0)[r]{\strut{} 0.6}}%
      \put(1078,3961){\makebox(0,0)[r]{\strut{} 0.8}}%
      \put(1078,4775){\makebox(0,0)[r]{\strut{} 1}}%
      \put(1210,484){\makebox(0,0){\strut{} 0}}%
      \put(1847,484){\makebox(0,0){\strut{} 2}}%
      \put(2484,484){\makebox(0,0){\strut{} 4}}%
      \put(3122,484){\makebox(0,0){\strut{} 6}}%
      \put(3759,484){\makebox(0,0){\strut{} 8}}%
      \put(4396,484){\makebox(0,0){\strut{} 10}}%
      \put(5033,484){\makebox(0,0){\strut{} 12}}%
      \put(5670,484){\makebox(0,0){\strut{} 14}}%
      \put(6121,704){\makebox(0,0)[l]{\strut{}-20}}%
      \put(6121,1213){\makebox(0,0)[l]{\strut{}-18}}%
      \put(6121,1722){\makebox(0,0)[l]{\strut{}-16}}%
      \put(6121,2231){\makebox(0,0)[l]{\strut{}-14}}%
      \put(6121,2740){\makebox(0,0)[l]{\strut{}-12}}%
      \put(6121,3248){\makebox(0,0)[l]{\strut{}-10}}%
      \put(6121,3757){\makebox(0,0)[l]{\strut{}-8}}%
      \put(6121,4266){\makebox(0,0)[l]{\strut{}-6}}%
      \put(6121,4775){\makebox(0,0)[l]{\strut{}-4}}%
      \put(308,2739){\rotatebox{-270}{\makebox(0,0){\strut{}$\omega_{t} / t$}}}%
      \put(6758,2739){\rotatebox{-270}{\makebox(0,0){\strut{}E / t}}}%
      \put(3599,154){\makebox(0,0){\strut{}$\#$ state}}%
    }%
    \gplgaddtomacro\gplfronttext{%
      \csname LTb\endcsname%
      \put(3222,3037){\makebox(0,0)[r]{\strut{}trionic weight U = 01}}%
      \csname LTb\endcsname%
      \put(3222,2817){\makebox(0,0)[r]{\strut{}trionic weight U = 06}}%
      \csname LTb\endcsname%
      \put(3222,2597){\makebox(0,0)[r]{\strut{}total energy U = 01}}%
      \csname LTb\endcsname%
      \put(3222,2377){\makebox(0,0)[r]{\strut{}total energy U = 06}}%
    }%
    \gplbacktext
    \put(0,0){\includegraphics{trionWeight-EiVal-14111}}%
    \gplfronttext
  \end{picture}%
\endgroup

%% file: img/trionCommut-EnPot-19111.tex
\begingroup
  \makeatletter
  \providecommand\color[2][]{%
    \GenericError{(gnuplot) \space\space\space\@spaces}{%
      Package color not loaded in conjunction with
      terminal option `colourtext'%
    }{See the gnuplot documentation for explanation.%
    }{Either use 'blacktext' in gnuplot or load the package
      color.sty in LaTeX.}%
    \renewcommand\color[2][]{}%
  }%
  \providecommand\includegraphics[2][]{%
    \GenericError{(gnuplot) \space\space\space\@spaces}{%
      Package graphicx or graphics not loaded%
    }{See the gnuplot documentation for explanation.%
    }{The gnuplot epslatex terminal needs graphicx.sty or graphics.sty.}%
    \renewcommand\includegraphics[2][]{}%
  }%
  \providecommand\rotatebox[2]{#2}%
  \@ifundefined{ifGPcolor}{%
    \newif\ifGPcolor
    \GPcolortrue
  }{}%
  \@ifundefined{ifGPblacktext}{%
    \newif\ifGPblacktext
    \GPblacktexttrue
  }{}%
  \let\gplgaddtomacro\g@addto@macro
  \gdef\gplbacktext{}%
  \gdef\gplfronttext{}%
  \makeatother
  \ifGPblacktext
    \def\colorrgb#1{}%
    \def\colorgray#1{}%
  \else
    \ifGPcolor
      \def\colorrgb#1{\color[rgb]{#1}}%
      \def\colorgray#1{\color[gray]{#1}}%
      \expandafter\def\csname LTw\endcsname{\color{white}}%
      \expandafter\def\csname LTb\endcsname{\color{black}}%
      \expandafter\def\csname LTa\endcsname{\color{black}}%
      \expandafter\def\csname LT0\endcsname{\color[rgb]{1,0,0}}%
      \expandafter\def\csname LT1\endcsname{\color[rgb]{0,1,0}}%
      \expandafter\def\csname LT2\endcsname{\color[rgb]{0,0,1}}%
      \expandafter\def\csname LT3\endcsname{\color[rgb]{1,0,1}}%
      \expandafter\def\csname LT4\endcsname{\color[rgb]{0,1,1}}%
      \expandafter\def\csname LT5\endcsname{\color[rgb]{1,1,0}}%
      \expandafter\def\csname LT6\endcsname{\color[rgb]{0,0,0}}%
      \expandafter\def\csname LT7\endcsname{\color[rgb]{1,0.3,0}}%
      \expandafter\def\csname LT8\endcsname{\color[rgb]{0.5,0.5,0.5}}%
    \else
      \def\colorrgb#1{\color{black}}%
      \def\colorgray#1{\color[gray]{#1}}%
      \expandafter\def\csname LTw\endcsname{\color{white}}%
      \expandafter\def\csname LTb\endcsname{\color{black}}%
      \expandafter\def\csname LTa\endcsname{\color{black}}%
      \expandafter\def\csname LT0\endcsname{\color{black}}%
      \expandafter\def\csname LT1\endcsname{\color{black}}%
      \expandafter\def\csname LT2\endcsname{\color{black}}%
      \expandafter\def\csname LT3\endcsname{\color{black}}%
      \expandafter\def\csname LT4\endcsname{\color{black}}%
      \expandafter\def\csname LT5\endcsname{\color{black}}%
      \expandafter\def\csname LT6\endcsname{\color{black}}%
      \expandafter\def\csname LT7\endcsname{\color{black}}%
      \expandafter\def\csname LT8\endcsname{\color{black}}%
    \fi
  \fi
  \setlength{\unitlength}{0.0500bp}%
  \begin{picture}(7200.00,5040.00)%
    \gplgaddtomacro\gplbacktext{%
      \csname LTb\endcsname%
      \put(1078,704){\makebox(0,0)[r]{\strut{} 0}}%
      \put(1078,1518){\makebox(0,0)[r]{\strut{} 0.2}}%
      \put(1078,2332){\makebox(0,0)[r]{\strut{} 0.4}}%
      \put(1078,3147){\makebox(0,0)[r]{\strut{} 0.6}}%
      \put(1078,3961){\makebox(0,0)[r]{\strut{} 0.8}}%
      \put(1078,4775){\makebox(0,0)[r]{\strut{} 1}}%
      \put(1210,484){\makebox(0,0){\strut{} 0}}%
      \put(2405,484){\makebox(0,0){\strut{} 5}}%
      \put(3599,484){\makebox(0,0){\strut{} 10}}%
      \put(4794,484){\makebox(0,0){\strut{} 15}}%
      \put(5989,484){\makebox(0,0){\strut{} 20}}%
      \put(6121,952){\makebox(0,0)[l]{\strut{}-12}}%
      \put(6121,1589){\makebox(0,0)[l]{\strut{}-10}}%
      \put(6121,2226){\makebox(0,0)[l]{\strut{}-8}}%
      \put(6121,2863){\makebox(0,0)[l]{\strut{}-6}}%
      \put(6121,3501){\makebox(0,0)[l]{\strut{}-4}}%
      \put(6121,4138){\makebox(0,0)[l]{\strut{}-2}}%
      \put(6121,4775){\makebox(0,0)[l]{\strut{} 0}}%
      \put(308,2739){\rotatebox{-270}{\makebox(0,0){\strut{}$[t^{\dagger},t]_{+}$}}}%
      \put(6758,2739){\rotatebox{-270}{\makebox(0,0){\strut{}E / t}}}%
      \put(3599,154){\makebox(0,0){\strut{}$\#$ state}}%
    }%
    \gplgaddtomacro\gplfronttext{%
      \csname LTb\endcsname%
      \put(3700,3037){\makebox(0,0)[r]{\strut{}trion anticommutator U = 01}}%
      \csname LTb\endcsname%
      \put(3700,2817){\makebox(0,0)[r]{\strut{}trion anticommutator U = 06}}%
      \csname LTb\endcsname%
      \put(3700,2597){\makebox(0,0)[r]{\strut{}interaction energy U = 01}}%
      \csname LTb\endcsname%
      \put(3700,2377){\makebox(0,0)[r]{\strut{}interaction energy U = 06}}%
    }%
    \gplbacktext
    \put(0,0){\includegraphics{trionCommut-EnPot-19111}}%
    \gplfronttext
  \end{picture}%
\endgroup

%% file: img/phasediagram-inset-14111.tex
\begingroup
  \makeatletter
  \providecommand\color[2][]{%
    \GenericError{(gnuplot) \space\space\space\@spaces}{%
      Package color not loaded in conjunction with
      terminal option `colourtext'%
    }{See the gnuplot documentation for explanation.%
    }{Either use 'blacktext' in gnuplot or load the package
      color.sty in LaTeX.}%
    \renewcommand\color[2][]{}%
  }%
  \providecommand\includegraphics[2][]{%
    \GenericError{(gnuplot) \space\space\space\@spaces}{%
      Package graphicx or graphics not loaded%
    }{See the gnuplot documentation for explanation.%
    }{The gnuplot epslatex terminal needs graphicx.sty or graphics.sty.}%
    \renewcommand\includegraphics[2][]{}%
  }%
  \providecommand\rotatebox[2]{#2}%
  \@ifundefined{ifGPcolor}{%
    \newif\ifGPcolor
    \GPcolortrue
  }{}%
  \@ifundefined{ifGPblacktext}{%
    \newif\ifGPblacktext
    \GPblacktexttrue
  }{}%
  \let\gplgaddtomacro\g@addto@macro
  \gdef\gplbacktext{}%
  \gdef\gplfronttext{}%
  \makeatother
  \ifGPblacktext
    \def\colorrgb#1{}%
    \def\colorgray#1{}%
  \else
    \ifGPcolor
      \def\colorrgb#1{\color[rgb]{#1}}%
      \def\colorgray#1{\color[gray]{#1}}%
      \expandafter\def\csname LTw\endcsname{\color{white}}%
      \expandafter\def\csname LTb\endcsname{\color{black}}%
      \expandafter\def\csname LTa\endcsname{\color{black}}%
      \expandafter\def\csname LT0\endcsname{\color[rgb]{1,0,0}}%
      \expandafter\def\csname LT1\endcsname{\color[rgb]{0,1,0}}%
      \expandafter\def\csname LT2\endcsname{\color[rgb]{0,0,1}}%
      \expandafter\def\csname LT3\endcsname{\color[rgb]{1,0,1}}%
      \expandafter\def\csname LT4\endcsname{\color[rgb]{0,1,1}}%
      \expandafter\def\csname LT5\endcsname{\color[rgb]{1,1,0}}%
      \expandafter\def\csname LT6\endcsname{\color[rgb]{0,0,0}}%
      \expandafter\def\csname LT7\endcsname{\color[rgb]{1,0.3,0}}%
      \expandafter\def\csname LT8\endcsname{\color[rgb]{0.5,0.5,0.5}}%
    \else
      \def\colorrgb#1{\color{black}}%
      \def\colorgray#1{\color[gray]{#1}}%
      \expandafter\def\csname LTw\endcsname{\color{white}}%
      \expandafter\def\csname LTb\endcsname{\color{black}}%
      \expandafter\def\csname LTa\endcsname{\color{black}}%
      \expandafter\def\csname LT0\endcsname{\color{black}}%
      \expandafter\def\csname LT1\endcsname{\color{black}}%
      \expandafter\def\csname LT2\endcsname{\color{black}}%
      \expandafter\def\csname LT3\endcsname{\color{black}}%
      \expandafter\def\csname LT4\endcsname{\color{black}}%
      \expandafter\def\csname LT5\endcsname{\color{black}}%
      \expandafter\def\csname LT6\endcsname{\color{black}}%
      \expandafter\def\csname LT7\endcsname{\color{black}}%
      \expandafter\def\csname LT8\endcsname{\color{black}}%
    \fi
  \fi
  \setlength{\unitlength}{0.0500bp}%
  \begin{picture}(7200.00,5040.00)%
    \gplgaddtomacro\gplbacktext{%
    }%
    \gplgaddtomacro\gplfronttext{%
      \csname LTb\endcsname%
      \put(1170,718){\makebox(0,0){\strut{} 0}}%
      \put(1777,718){\makebox(0,0){\strut{} 1}}%
      \put(2385,718){\makebox(0,0){\strut{} 2}}%
      \put(2993,718){\makebox(0,0){\strut{} 3}}%
      \put(3600,718){\makebox(0,0){\strut{} 4}}%
      \put(4207,718){\makebox(0,0){\strut{} 5}}%
      \put(4815,718){\makebox(0,0){\strut{} 6}}%
      \put(5423,718){\makebox(0,0){\strut{} 7}}%
      \put(6030,718){\makebox(0,0){\strut{} 8}}%
      \put(3600,388){\makebox(0,0){\strut{}$ U_{12}/t = U_{13}/t  = U/t $ }}%
      \put(916,1086){\makebox(0,0)[r]{\strut{}-2}}%
      \put(916,1704){\makebox(0,0)[r]{\strut{} 0}}%
      \put(916,2322){\makebox(0,0)[r]{\strut{} 2}}%
      \put(916,2938){\makebox(0,0)[r]{\strut{} 4}}%
      \put(916,3556){\makebox(0,0)[r]{\strut{} 6}}%
      \put(916,4174){\makebox(0,0)[r]{\strut{} 8}}%
      \put(586,2630){\rotatebox{-270}{\makebox(0,0){\strut{}$ U_{23}/t $}}}%
      \put(6653,1085){\makebox(0,0)[l]{\strut{} 1e-09}}%
      \put(6653,1428){\makebox(0,0)[l]{\strut{} 1e-08}}%
      \put(6653,1771){\makebox(0,0)[l]{\strut{} 1e-07}}%
      \put(6653,2114){\makebox(0,0)[l]{\strut{} 1e-06}}%
      \put(6653,2457){\makebox(0,0)[l]{\strut{} 1e-05}}%
      \put(6653,2801){\makebox(0,0)[l]{\strut{} 0.0001}}%
      \put(6653,3144){\makebox(0,0)[l]{\strut{} 0.001}}%
      \put(6653,3487){\makebox(0,0)[l]{\strut{} 0.01}}%
      \put(6653,3830){\makebox(0,0)[l]{\strut{} 0.1}}%
      \put(6653,4174){\makebox(0,0)[l]{\strut{} 1}}%
    }%
    \gplgaddtomacro\gplfronttext{%
      \put(3478,2279){\makebox(0,0)[r]{\strut{}-2}}%
      \put(3478,2658){\makebox(0,0)[r]{\strut{} 0}}%
      \put(3478,3036){\makebox(0,0)[r]{\strut{} 2}}%
      \put(3478,3415){\makebox(0,0)[r]{\strut{} 4}}%
      \put(3478,3793){\makebox(0,0)[r]{\strut{} 6}}%
      \put(3478,4172){\makebox(0,0)[r]{\strut{} 8}}%
      \put(3673,1996){\makebox(0,0){\strut{} 0}}%
      \put(3967,1996){\makebox(0,0){\strut{} 0.5}}%
      \put(4261,1996){\makebox(0,0){\strut{} 1}}%
      \put(4555,1996){\makebox(0,0){\strut{} 1.5}}%
      \put(4850,1996){\makebox(0,0){\strut{} 2}}%
      \put(5144,1996){\makebox(0,0){\strut{} 2.5}}%
      \put(5438,1996){\makebox(0,0){\strut{} 3}}%
      \put(5732,1996){\makebox(0,0){\strut{} 3.5}}%
      \put(6026,1996){\makebox(0,0){\strut{} 4}}%
      \put(2972,3225){\rotatebox{-270}{\makebox(0,0){\strut{}$ E_{gap} / t $}}}%
      \put(4849,1666){\makebox(0,0){\strut{}$ | U | / t $}}%
      \put(4261,3793){\makebox(0,0)[l]{\strut{}trionic regime}}%
    }%
    \gplgaddtomacro\gplfronttext{%
    }%
    \gplbacktext
    \put(0,0){\includegraphics{phasediagram-inset-14111}}%
    \gplfronttext
  \end{picture}%
\endgroup

%% file: img/trionVSboson.tex
\begingroup
  \makeatletter
  \providecommand\color[2][]{%
    \GenericError{(gnuplot) \space\space\space\@spaces}{%
      Package color not loaded in conjunction with
      terminal option `colourtext'%
    }{See the gnuplot documentation for explanation.%
    }{Either use 'blacktext' in gnuplot or load the package
      color.sty in LaTeX.}%
    \renewcommand\color[2][]{}%
  }%
  \providecommand\includegraphics[2][]{%
    \GenericError{(gnuplot) \space\space\space\@spaces}{%
      Package graphicx or graphics not loaded%
    }{See the gnuplot documentation for explanation.%
    }{The gnuplot epslatex terminal needs graphicx.sty or graphics.sty.}%
    \renewcommand\includegraphics[2][]{}%
  }%
  \providecommand\rotatebox[2]{#2}%
  \@ifundefined{ifGPcolor}{%
    \newif\ifGPcolor
    \GPcolortrue
  }{}%
  \@ifundefined{ifGPblacktext}{%
    \newif\ifGPblacktext
    \GPblacktexttrue
  }{}%
  \let\gplgaddtomacro\g@addto@macro
  \gdef\gplbacktext{}%
  \gdef\gplfronttext{}%
  \makeatother
  \ifGPblacktext
    \def\colorrgb#1{}%
    \def\colorgray#1{}%
  \else
    \ifGPcolor
      \def\colorrgb#1{\color[rgb]{#1}}%
      \def\colorgray#1{\color[gray]{#1}}%
      \expandafter\def\csname LTw\endcsname{\color{white}}%
      \expandafter\def\csname LTb\endcsname{\color{black}}%
      \expandafter\def\csname LTa\endcsname{\color{black}}%
      \expandafter\def\csname LT0\endcsname{\color[rgb]{1,0,0}}%
      \expandafter\def\csname LT1\endcsname{\color[rgb]{0,1,0}}%
      \expandafter\def\csname LT2\endcsname{\color[rgb]{0,0,1}}%
      \expandafter\def\csname LT3\endcsname{\color[rgb]{1,0,1}}%
      \expandafter\def\csname LT4\endcsname{\color[rgb]{0,1,1}}%
      \expandafter\def\csname LT5\endcsname{\color[rgb]{1,1,0}}%
      \expandafter\def\csname LT6\endcsname{\color[rgb]{0,0,0}}%
      \expandafter\def\csname LT7\endcsname{\color[rgb]{1,0.3,0}}%
      \expandafter\def\csname LT8\endcsname{\color[rgb]{0.5,0.5,0.5}}%
    \else
      \def\colorrgb#1{\color{black}}%
      \def\colorgray#1{\color[gray]{#1}}%
      \expandafter\def\csname LTw\endcsname{\color{white}}%
      \expandafter\def\csname LTb\endcsname{\color{black}}%
      \expandafter\def\csname LTa\endcsname{\color{black}}%
      \expandafter\def\csname LT0\endcsname{\color{black}}%
      \expandafter\def\csname LT1\endcsname{\color{black}}%
      \expandafter\def\csname LT2\endcsname{\color{black}}%
      \expandafter\def\csname LT3\endcsname{\color{black}}%
      \expandafter\def\csname LT4\endcsname{\color{black}}%
      \expandafter\def\csname LT5\endcsname{\color{black}}%
      \expandafter\def\csname LT6\endcsname{\color{black}}%
      \expandafter\def\csname LT7\endcsname{\color{black}}%
      \expandafter\def\csname LT8\endcsname{\color{black}}%
    \fi
  \fi
  \setlength{\unitlength}{0.0500bp}%
  \begin{picture}(7200.00,5040.00)%
    \gplgaddtomacro\gplbacktext{%
      \csname LTb\endcsname%
      \put(858,704){\makebox(0,0)[r]{\strut{} 0}}%
      \put(858,1111){\makebox(0,0)[r]{\strut{} 0.1}}%
      \put(858,1518){\makebox(0,0)[r]{\strut{} 0.2}}%
      \put(858,1925){\makebox(0,0)[r]{\strut{} 0.3}}%
      \put(858,2332){\makebox(0,0)[r]{\strut{} 0.4}}%
      \put(858,2740){\makebox(0,0)[r]{\strut{} 0.5}}%
      \put(858,3147){\makebox(0,0)[r]{\strut{} 0.6}}%
      \put(858,3554){\makebox(0,0)[r]{\strut{} 0.7}}%
      \put(858,3961){\makebox(0,0)[r]{\strut{} 0.8}}%
      \put(858,4368){\makebox(0,0)[r]{\strut{} 0.9}}%
      \put(858,4775){\makebox(0,0)[r]{\strut{} 1}}%
      \put(990,484){\makebox(0,0){\strut{} 0}}%
      \put(1643,484){\makebox(0,0){\strut{} 0.5}}%
      \put(2296,484){\makebox(0,0){\strut{} 1}}%
      \put(2950,484){\makebox(0,0){\strut{} 1.5}}%
      \put(3603,484){\makebox(0,0){\strut{} 2}}%
      \put(4256,484){\makebox(0,0){\strut{} 2.5}}%
      \put(4909,484){\makebox(0,0){\strut{} 3}}%
      \put(5563,484){\makebox(0,0){\strut{} 3.5}}%
      \put(6216,484){\makebox(0,0){\strut{} 4}}%
      \put(6869,484){\makebox(0,0){\strut{} 4.5}}%
      \put(3929,154){\makebox(0,0){\strut{}U / t}}%
    }%
    \gplgaddtomacro\gplfronttext{%
      \csname LTb\endcsname%
      \put(2839,4665){\makebox(0,0)[r]{\strut{}trion 10 sites}}%
      \csname LTb\endcsname%
      \put(2839,4445){\makebox(0,0)[r]{\strut{}trion 20 sites}}%
      \csname LTb\endcsname%
      \put(2839,4225){\makebox(0,0)[r]{\strut{}pair 10 sites}}%
      \csname LTb\endcsname%
      \put(2839,4005){\makebox(0,0)[r]{\strut{}pair 10 sites}}%
    }%
    \gplbacktext
    \put(0,0){\includegraphics{trionVSboson}}%
    \gplfronttext
  \end{picture}%
\endgroup

%% file: img/1D-pairing-12333-final.tex
\begingroup
  \makeatletter
  \providecommand\color[2][]{%
    \GenericError{(gnuplot) \space\space\space\@spaces}{%
      Package color not loaded in conjunction with
      terminal option `colourtext'%
    }{See the gnuplot documentation for explanation.%
    }{Either use 'blacktext' in gnuplot or load the package
      color.sty in LaTeX.}%
    \renewcommand\color[2][]{}%
  }%
  \providecommand\includegraphics[2][]{%
    \GenericError{(gnuplot) \space\space\space\@spaces}{%
      Package graphicx or graphics not loaded%
    }{See the gnuplot documentation for explanation.%
    }{The gnuplot epslatex terminal needs graphicx.sty or graphics.sty.}%
    \renewcommand\includegraphics[2][]{}%
  }%
  \providecommand\rotatebox[2]{#2}%
  \@ifundefined{ifGPcolor}{%
    \newif\ifGPcolor
    \GPcolortrue
  }{}%
  \@ifundefined{ifGPblacktext}{%
    \newif\ifGPblacktext
    \GPblacktexttrue
  }{}%
  \let\gplgaddtomacro\g@addto@macro
  \gdef\gplbacktext{}%
  \gdef\gplfronttext{}%
  \makeatother
  \ifGPblacktext
    \def\colorrgb#1{}%
    \def\colorgray#1{}%
  \else
    \ifGPcolor
      \def\colorrgb#1{\color[rgb]{#1}}%
      \def\colorgray#1{\color[gray]{#1}}%
      \expandafter\def\csname LTw\endcsname{\color{white}}%
      \expandafter\def\csname LTb\endcsname{\color{black}}%
      \expandafter\def\csname LTa\endcsname{\color{black}}%
      \expandafter\def\csname LT0\endcsname{\color[rgb]{1,0,0}}%
      \expandafter\def\csname LT1\endcsname{\color[rgb]{0,1,0}}%
      \expandafter\def\csname LT2\endcsname{\color[rgb]{0,0,1}}%
      \expandafter\def\csname LT3\endcsname{\color[rgb]{1,0,1}}%
      \expandafter\def\csname LT4\endcsname{\color[rgb]{0,1,1}}%
      \expandafter\def\csname LT5\endcsname{\color[rgb]{1,1,0}}%
      \expandafter\def\csname LT6\endcsname{\color[rgb]{0,0,0}}%
      \expandafter\def\csname LT7\endcsname{\color[rgb]{1,0.3,0}}%
      \expandafter\def\csname LT8\endcsname{\color[rgb]{0.5,0.5,0.5}}%
    \else
      \def\colorrgb#1{\color{black}}%
      \def\colorgray#1{\color[gray]{#1}}%
      \expandafter\def\csname LTw\endcsname{\color{white}}%
      \expandafter\def\csname LTb\endcsname{\color{black}}%
      \expandafter\def\csname LTa\endcsname{\color{black}}%
      \expandafter\def\csname LT0\endcsname{\color{black}}%
      \expandafter\def\csname LT1\endcsname{\color{black}}%
      \expandafter\def\csname LT2\endcsname{\color{black}}%
      \expandafter\def\csname LT3\endcsname{\color{black}}%
      \expandafter\def\csname LT4\endcsname{\color{black}}%
      \expandafter\def\csname LT5\endcsname{\color{black}}%
      \expandafter\def\csname LT6\endcsname{\color{black}}%
      \expandafter\def\csname LT7\endcsname{\color{black}}%
      \expandafter\def\csname LT8\endcsname{\color{black}}%
    \fi
  \fi
  \setlength{\unitlength}{0.0500bp}%
  \begin{picture}(7200.00,5040.00)%
    \gplgaddtomacro\gplbacktext{%
    }%
    \gplgaddtomacro\gplfronttext{%
      \csname LTb\endcsname%
      \put(3267,718){\makebox(0,0){\strut{} 0}}%
      \put(3876,718){\makebox(0,0){\strut{} 1}}%
      \put(4484,718){\makebox(0,0){\strut{} 2}}%
      \put(5092,718){\makebox(0,0){\strut{} 3}}%
      \put(5700,718){\makebox(0,0){\strut{} 4}}%
      \put(6309,718){\makebox(0,0){\strut{} 5}}%
      \put(4788,388){\makebox(0,0){\strut{}$U/t$}}%
      \put(6669,1085){\makebox(0,0)[l]{\strut{}$10^{-4}$}}%
      \put(6669,1702){\makebox(0,0)[l]{\strut{}$10^{-3}$}}%
      \put(6669,2320){\makebox(0,0)[l]{\strut{}$10^{-2}$}}%
      \put(6669,2938){\makebox(0,0)[l]{\strut{}$10^{-1}$}}%
      \put(6669,3556){\makebox(0,0)[l]{\strut{}$10^{0}$}}%
      \put(6669,4174){\makebox(0,0)[l]{\strut{}$10^{1}$}}%
    }%
    \gplgaddtomacro\gplbacktext{%
      \put(462,4167){\makebox(0,0)[r]{\strut{} 1}}%
      \put(462,3550){\makebox(0,0)[r]{\strut{} 2}}%
      \put(462,2934){\makebox(0,0)[r]{\strut{} 3}}%
      \put(462,2317){\makebox(0,0)[r]{\strut{} 4}}%
      \put(462,1701){\makebox(0,0)[r]{\strut{} 5}}%
      \put(462,1084){\makebox(0,0)[r]{\strut{} 6}}%
      \put(657,801){\makebox(0,0){\strut{}$10^{-5}$}}%
      \put(1179,801){\makebox(0,0){\strut{}$10^{-4}$}}%
      \put(1702,801){\makebox(0,0){\strut{}$10^{-3}$}}%
      \put(2224,801){\makebox(0,0){\strut{}$10^{-2}$}}%
      \put(2747,801){\makebox(0,0){\strut{}$10^{-1}$}}%
      \put(3269,801){\makebox(0,0){\strut{}$10^{0}$}}%
      \put(-44,2625){\rotatebox{-270}{\makebox(0,0){\strut{}$d_{x}$}}}%
      \put(1963,471){\makebox(0,0){\strut{}Pairing correlation}}%
    }%
    \gplgaddtomacro\gplfronttext{%
      \csname LTb\endcsname%
      \put(1245,4057){\makebox(0,0)[l]{\strut{}$U/t=0.01$}}%
      \csname LTb\endcsname%
      \put(1245,3837){\makebox(0,0)[l]{\strut{}$U/t=1.60$}}%
      \csname LTb\endcsname%
      \put(1245,3617){\makebox(0,0)[l]{\strut{}$U/t=6.00$}}%
    }%
    \gplgaddtomacro\gplbacktext{%
      \csname LTb\endcsname%
      \put(6440,4169){\makebox(0,0)[l]{\strut{} 0}}%
      \put(6440,4472){\makebox(0,0)[l]{\strut{} 1}}%
      \put(6440,4775){\makebox(0,0)[l]{\strut{} 2}}%
      \put(6945,4472){\rotatebox{-270}{\makebox(0,0){\strut{}$\lambda, f$}}}%
    }%
    \gplgaddtomacro\gplfronttext{%
      \csname LTb\endcsname%
      \put(3860,4710){\makebox(0,0)[l]{\strut{}$f$}}%
      \csname LTb\endcsname%
      \put(3860,4579){\makebox(0,0)[l]{\strut{}$\lambda$}}%
    }%
    \gplbacktext
    \put(0,0){\includegraphics{1D-pairing-12333-final}}%
    \gplfronttext
  \end{picture}%
\endgroup

%% file: img/trionic-green-function-N06-111vs222-U04.tex
\begingroup
  \makeatletter
  \providecommand\color[2][]{%
    \GenericError{(gnuplot) \space\space\space\@spaces}{%
      Package color not loaded in conjunction with
      terminal option `colourtext'%
    }{See the gnuplot documentation for explanation.%
    }{Either use 'blacktext' in gnuplot or load the package
      color.sty in LaTeX.}%
    \renewcommand\color[2][]{}%
  }%
  \providecommand\includegraphics[2][]{%
    \GenericError{(gnuplot) \space\space\space\@spaces}{%
      Package graphicx or graphics not loaded%
    }{See the gnuplot documentation for explanation.%
    }{The gnuplot epslatex terminal needs graphicx.sty or graphics.sty.}%
    \renewcommand\includegraphics[2][]{}%
  }%
  \providecommand\rotatebox[2]{#2}%
  \@ifundefined{ifGPcolor}{%
    \newif\ifGPcolor
    \GPcolortrue
  }{}%
  \@ifundefined{ifGPblacktext}{%
    \newif\ifGPblacktext
    \GPblacktexttrue
  }{}%
  \let\gplgaddtomacro\g@addto@macro
  \gdef\gplbacktext{}%
  \gdef\gplfronttext{}%
  \makeatother
  \ifGPblacktext
    \def\colorrgb#1{}%
    \def\colorgray#1{}%
  \else
    \ifGPcolor
      \def\colorrgb#1{\color[rgb]{#1}}%
      \def\colorgray#1{\color[gray]{#1}}%
      \expandafter\def\csname LTw\endcsname{\color{white}}%
      \expandafter\def\csname LTb\endcsname{\color{black}}%
      \expandafter\def\csname LTa\endcsname{\color{black}}%
      \expandafter\def\csname LT0\endcsname{\color[rgb]{1,0,0}}%
      \expandafter\def\csname LT1\endcsname{\color[rgb]{0,1,0}}%
      \expandafter\def\csname LT2\endcsname{\color[rgb]{0,0,1}}%
      \expandafter\def\csname LT3\endcsname{\color[rgb]{1,0,1}}%
      \expandafter\def\csname LT4\endcsname{\color[rgb]{0,1,1}}%
      \expandafter\def\csname LT5\endcsname{\color[rgb]{1,1,0}}%
      \expandafter\def\csname LT6\endcsname{\color[rgb]{0,0,0}}%
      \expandafter\def\csname LT7\endcsname{\color[rgb]{1,0.3,0}}%
      \expandafter\def\csname LT8\endcsname{\color[rgb]{0.5,0.5,0.5}}%
    \else
      \def\colorrgb#1{\color{black}}%
      \def\colorgray#1{\color[gray]{#1}}%
      \expandafter\def\csname LTw\endcsname{\color{white}}%
      \expandafter\def\csname LTb\endcsname{\color{black}}%
      \expandafter\def\csname LTa\endcsname{\color{black}}%
      \expandafter\def\csname LT0\endcsname{\color{black}}%
      \expandafter\def\csname LT1\endcsname{\color{black}}%
      \expandafter\def\csname LT2\endcsname{\color{black}}%
      \expandafter\def\csname LT3\endcsname{\color{black}}%
      \expandafter\def\csname LT4\endcsname{\color{black}}%
      \expandafter\def\csname LT5\endcsname{\color{black}}%
      \expandafter\def\csname LT6\endcsname{\color{black}}%
      \expandafter\def\csname LT7\endcsname{\color{black}}%
      \expandafter\def\csname LT8\endcsname{\color{black}}%
    \fi
  \fi
  \setlength{\unitlength}{0.0500bp}%
  \begin{picture}(7200.00,5040.00)%
    \gplgaddtomacro\gplbacktext{%
    }%
    \gplgaddtomacro\gplfronttext{%
      \csname LTb\endcsname%
      \put(916,2894){\makebox(0,0)[r]{\strut{}-13}}%
      \put(916,3241){\makebox(0,0)[r]{\strut{}-12.5}}%
      \put(916,3587){\makebox(0,0)[r]{\strut{}-12}}%
      \put(916,3933){\makebox(0,0)[r]{\strut{}-11.5}}%
      \put(916,4280){\makebox(0,0)[r]{\strut{}-11}}%
      \put(272,3587){\rotatebox{-270}{\makebox(0,0){\strut{}$ \omega/t $}}}%
      \put(6653,2753){\makebox(0,0)[l]{\strut{} 0.01}}%
      \put(6653,3170){\makebox(0,0)[l]{\strut{} 0.1}}%
      \put(6653,3586){\makebox(0,0)[l]{\strut{} 1}}%
      \put(6653,4002){\makebox(0,0)[l]{\strut{} 10}}%
      \put(6653,4419){\makebox(0,0)[l]{\strut{} 100}}%
    }%
    \gplgaddtomacro\gplbacktext{%
    }%
    \gplgaddtomacro\gplfronttext{%
      \csname LTb\endcsname%
      \put(1575,472){\makebox(0,0){\strut{}$ + \pi $}}%
      \put(2385,472){\makebox(0,0){\strut{}$ + \frac{2\pi}{3} $}}%
      \put(3195,472){\makebox(0,0){\strut{}$ + \frac{\pi}{3} $}}%
      \put(4005,472){\makebox(0,0){\strut{}0}}%
      \put(4815,472){\makebox(0,0){\strut{}$ - \frac{\pi}{3} $}}%
      \put(5625,472){\makebox(0,0){\strut{}$ - \frac{2\pi}{3} $}}%
      \put(3600,142){\makebox(0,0){\strut{}$ k $}}%
      \put(916,979){\makebox(0,0)[r]{\strut{}-13}}%
      \put(916,1326){\makebox(0,0)[r]{\strut{}-12.5}}%
      \put(916,1672){\makebox(0,0)[r]{\strut{}-12}}%
      \put(916,2018){\makebox(0,0)[r]{\strut{}-11.5}}%
      \put(916,2365){\makebox(0,0)[r]{\strut{}-11}}%
      \put(190,1672){\rotatebox{-270}{\makebox(0,0){\strut{}$ \omega/t $}}}%
      \put(6653,838){\makebox(0,0)[l]{\strut{} 0.01}}%
      \put(6653,1255){\makebox(0,0)[l]{\strut{} 0.1}}%
      \put(6653,1671){\makebox(0,0)[l]{\strut{} 1}}%
      \put(6653,2087){\makebox(0,0)[l]{\strut{} 10}}%
      \put(6653,2504){\makebox(0,0)[l]{\strut{} 100}}%
    }%
    \gplbacktext
    \put(0,0){\includegraphics{trionic-green-function-N06-111vs222-U04}}%
    \gplfronttext
  \end{picture}%
\endgroup

%% file: img/trionic-green-function-N06-111vs222-U08.tex
\begingroup
  \makeatletter
  \providecommand\color[2][]{%
    \GenericError{(gnuplot) \space\space\space\@spaces}{%
      Package color not loaded in conjunction with
      terminal option `colourtext'%
    }{See the gnuplot documentation for explanation.%
    }{Either use 'blacktext' in gnuplot or load the package
      color.sty in LaTeX.}%
    \renewcommand\color[2][]{}%
  }%
  \providecommand\includegraphics[2][]{%
    \GenericError{(gnuplot) \space\space\space\@spaces}{%
      Package graphicx or graphics not loaded%
    }{See the gnuplot documentation for explanation.%
    }{The gnuplot epslatex terminal needs graphicx.sty or graphics.sty.}%
    \renewcommand\includegraphics[2][]{}%
  }%
  \providecommand\rotatebox[2]{#2}%
  \@ifundefined{ifGPcolor}{%
    \newif\ifGPcolor
    \GPcolortrue
  }{}%
  \@ifundefined{ifGPblacktext}{%
    \newif\ifGPblacktext
    \GPblacktexttrue
  }{}%
  \let\gplgaddtomacro\g@addto@macro
  \gdef\gplbacktext{}%
  \gdef\gplfronttext{}%
  \makeatother
  \ifGPblacktext
    \def\colorrgb#1{}%
    \def\colorgray#1{}%
  \else
    \ifGPcolor
      \def\colorrgb#1{\color[rgb]{#1}}%
      \def\colorgray#1{\color[gray]{#1}}%
      \expandafter\def\csname LTw\endcsname{\color{white}}%
      \expandafter\def\csname LTb\endcsname{\color{black}}%
      \expandafter\def\csname LTa\endcsname{\color{black}}%
      \expandafter\def\csname LT0\endcsname{\color[rgb]{1,0,0}}%
      \expandafter\def\csname LT1\endcsname{\color[rgb]{0,1,0}}%
      \expandafter\def\csname LT2\endcsname{\color[rgb]{0,0,1}}%
      \expandafter\def\csname LT3\endcsname{\color[rgb]{1,0,1}}%
      \expandafter\def\csname LT4\endcsname{\color[rgb]{0,1,1}}%
      \expandafter\def\csname LT5\endcsname{\color[rgb]{1,1,0}}%
      \expandafter\def\csname LT6\endcsname{\color[rgb]{0,0,0}}%
      \expandafter\def\csname LT7\endcsname{\color[rgb]{1,0.3,0}}%
      \expandafter\def\csname LT8\endcsname{\color[rgb]{0.5,0.5,0.5}}%
    \else
      \def\colorrgb#1{\color{black}}%
      \def\colorgray#1{\color[gray]{#1}}%
      \expandafter\def\csname LTw\endcsname{\color{white}}%
      \expandafter\def\csname LTb\endcsname{\color{black}}%
      \expandafter\def\csname LTa\endcsname{\color{black}}%
      \expandafter\def\csname LT0\endcsname{\color{black}}%
      \expandafter\def\csname LT1\endcsname{\color{black}}%
      \expandafter\def\csname LT2\endcsname{\color{black}}%
      \expandafter\def\csname LT3\endcsname{\color{black}}%
      \expandafter\def\csname LT4\endcsname{\color{black}}%
      \expandafter\def\csname LT5\endcsname{\color{black}}%
      \expandafter\def\csname LT6\endcsname{\color{black}}%
      \expandafter\def\csname LT7\endcsname{\color{black}}%
      \expandafter\def\csname LT8\endcsname{\color{black}}%
    \fi
  \fi
  \setlength{\unitlength}{0.0500bp}%
  \begin{picture}(7200.00,5040.00)%
    \gplgaddtomacro\gplbacktext{%
    }%
    \gplgaddtomacro\gplfronttext{%
      \csname LTb\endcsname%
      \put(916,2894){\makebox(0,0)[r]{\strut{}-25}}%
      \put(916,3241){\makebox(0,0)[r]{\strut{}-24.5}}%
      \put(916,3587){\makebox(0,0)[r]{\strut{}-24}}%
      \put(916,3933){\makebox(0,0)[r]{\strut{}-23.5}}%
      \put(916,4280){\makebox(0,0)[r]{\strut{}-23}}%
      \put(272,3587){\rotatebox{-270}{\makebox(0,0){\strut{}$ \omega/t $}}}%
      \put(6653,2753){\makebox(0,0)[l]{\strut{} 0.01}}%
      \put(6653,3170){\makebox(0,0)[l]{\strut{} 0.1}}%
      \put(6653,3586){\makebox(0,0)[l]{\strut{} 1}}%
      \put(6653,4002){\makebox(0,0)[l]{\strut{} 10}}%
      \put(6653,4419){\makebox(0,0)[l]{\strut{} 100}}%
    }%
    \gplgaddtomacro\gplbacktext{%
    }%
    \gplgaddtomacro\gplfronttext{%
      \csname LTb\endcsname%
      \put(1575,472){\makebox(0,0){\strut{}$ + \pi $}}%
      \put(2385,472){\makebox(0,0){\strut{}$ + \frac{2\pi}{3} $}}%
      \put(3195,472){\makebox(0,0){\strut{}$ + \frac{\pi}{3} $}}%
      \put(4005,472){\makebox(0,0){\strut{}0}}%
      \put(4815,472){\makebox(0,0){\strut{}$ - \frac{\pi}{3} $}}%
      \put(5625,472){\makebox(0,0){\strut{}$ - \frac{2\pi}{3} $}}%
      \put(3600,142){\makebox(0,0){\strut{}$ k $}}%
      \put(916,979){\makebox(0,0)[r]{\strut{}-25}}%
      \put(916,1326){\makebox(0,0)[r]{\strut{}-24.5}}%
      \put(916,1672){\makebox(0,0)[r]{\strut{}-24}}%
      \put(916,2018){\makebox(0,0)[r]{\strut{}-23.5}}%
      \put(916,2365){\makebox(0,0)[r]{\strut{}-23}}%
      \put(190,1672){\rotatebox{-270}{\makebox(0,0){\strut{}$ \omega/t $}}}%
      \put(6653,838){\makebox(0,0)[l]{\strut{} 0.01}}%
      \put(6653,1255){\makebox(0,0)[l]{\strut{} 0.1}}%
      \put(6653,1671){\makebox(0,0)[l]{\strut{} 1}}%
      \put(6653,2087){\makebox(0,0)[l]{\strut{} 10}}%
      \put(6653,2504){\makebox(0,0)[l]{\strut{} 100}}%
    }%
    \gplbacktext
    \put(0,0){\includegraphics{trionic-green-function-N06-111vs222-U08}}%
    \gplfronttext
  \end{picture}%
\endgroup

%% file: img/effective-parameter-1Dvs2D.tex
\begingroup
  \makeatletter
  \providecommand\color[2][]{%
    \GenericError{(gnuplot) \space\space\space\@spaces}{%
      Package color not loaded in conjunction with
      terminal option `colourtext'%
    }{See the gnuplot documentation for explanation.%
    }{Either use 'blacktext' in gnuplot or load the package
      color.sty in LaTeX.}%
    \renewcommand\color[2][]{}%
  }%
  \providecommand\includegraphics[2][]{%
    \GenericError{(gnuplot) \space\space\space\@spaces}{%
      Package graphicx or graphics not loaded%
    }{See the gnuplot documentation for explanation.%
    }{The gnuplot epslatex terminal needs graphicx.sty or graphics.sty.}%
    \renewcommand\includegraphics[2][]{}%
  }%
  \providecommand\rotatebox[2]{#2}%
  \@ifundefined{ifGPcolor}{%
    \newif\ifGPcolor
    \GPcolortrue
  }{}%
  \@ifundefined{ifGPblacktext}{%
    \newif\ifGPblacktext
    \GPblacktexttrue
  }{}%
  \let\gplgaddtomacro\g@addto@macro
  \gdef\gplbacktext{}%
  \gdef\gplfronttext{}%
  \makeatother
  \ifGPblacktext
    \def\colorrgb#1{}%
    \def\colorgray#1{}%
  \else
    \ifGPcolor
      \def\colorrgb#1{\color[rgb]{#1}}%
      \def\colorgray#1{\color[gray]{#1}}%
      \expandafter\def\csname LTw\endcsname{\color{white}}%
      \expandafter\def\csname LTb\endcsname{\color{black}}%
      \expandafter\def\csname LTa\endcsname{\color{black}}%
      \expandafter\def\csname LT0\endcsname{\color[rgb]{1,0,0}}%
      \expandafter\def\csname LT1\endcsname{\color[rgb]{0,1,0}}%
      \expandafter\def\csname LT2\endcsname{\color[rgb]{0,0,1}}%
      \expandafter\def\csname LT3\endcsname{\color[rgb]{1,0,1}}%
      \expandafter\def\csname LT4\endcsname{\color[rgb]{0,1,1}}%
      \expandafter\def\csname LT5\endcsname{\color[rgb]{1,1,0}}%
      \expandafter\def\csname LT6\endcsname{\color[rgb]{0,0,0}}%
      \expandafter\def\csname LT7\endcsname{\color[rgb]{1,0.3,0}}%
      \expandafter\def\csname LT8\endcsname{\color[rgb]{0.5,0.5,0.5}}%
    \else
      \def\colorrgb#1{\color{black}}%
      \def\colorgray#1{\color[gray]{#1}}%
      \expandafter\def\csname LTw\endcsname{\color{white}}%
      \expandafter\def\csname LTb\endcsname{\color{black}}%
      \expandafter\def\csname LTa\endcsname{\color{black}}%
      \expandafter\def\csname LT0\endcsname{\color{black}}%
      \expandafter\def\csname LT1\endcsname{\color{black}}%
      \expandafter\def\csname LT2\endcsname{\color{black}}%
      \expandafter\def\csname LT3\endcsname{\color{black}}%
      \expandafter\def\csname LT4\endcsname{\color{black}}%
      \expandafter\def\csname LT5\endcsname{\color{black}}%
      \expandafter\def\csname LT6\endcsname{\color{black}}%
      \expandafter\def\csname LT7\endcsname{\color{black}}%
      \expandafter\def\csname LT8\endcsname{\color{black}}%
    \fi
  \fi
  \setlength{\unitlength}{0.0500bp}%
  \begin{picture}(7200.00,5040.00)%
    \gplgaddtomacro\gplbacktext{%
      \csname LTb\endcsname%
      \put(1078,767){\makebox(0,0)[r]{\strut{} 0}}%
      \put(1078,1369){\makebox(0,0)[r]{\strut{} 0.5}}%
      \put(1078,1971){\makebox(0,0)[r]{\strut{} 1}}%
      \put(1078,2573){\makebox(0,0)[r]{\strut{} 1.5}}%
      \put(1078,3175){\makebox(0,0)[r]{\strut{} 2}}%
      \put(1078,3777){\makebox(0,0)[r]{\strut{} 2.5}}%
      \put(1078,4379){\makebox(0,0)[r]{\strut{} 3}}%
      \put(1273,484){\makebox(0,0){\strut{} 0}}%
      \put(2403,484){\makebox(0,0){\strut{} 5}}%
      \put(3534,484){\makebox(0,0){\strut{} 10}}%
      \put(4664,484){\makebox(0,0){\strut{} 15}}%
      \put(5794,484){\makebox(0,0){\strut{} 20}}%
      \put(5989,767){\makebox(0,0)[l]{\strut{} 0}}%
      \put(5989,1128){\makebox(0,0)[l]{\strut{} 0.2}}%
      \put(5989,1489){\makebox(0,0)[l]{\strut{} 0.4}}%
      \put(5989,1851){\makebox(0,0)[l]{\strut{} 0.6}}%
      \put(5989,2212){\makebox(0,0)[l]{\strut{} 0.8}}%
      \put(5989,2573){\makebox(0,0)[l]{\strut{} 1}}%
      \put(5989,2934){\makebox(0,0)[l]{\strut{} 1.2}}%
      \put(5989,3295){\makebox(0,0)[l]{\strut{} 1.4}}%
      \put(5989,3657){\makebox(0,0)[l]{\strut{} 1.6}}%
      \put(5989,4018){\makebox(0,0)[l]{\strut{} 1.8}}%
      \put(5989,4379){\makebox(0,0)[l]{\strut{} 2}}%
      \put(308,2573){\rotatebox{-270}{\makebox(0,0){\strut{}$t_{eff}$}}}%
      \put(6758,2573){\rotatebox{-270}{\makebox(0,0){\strut{}$V_{eff}$}}}%
      \put(3533,154){\makebox(0,0){\strut{}$U/t$}}%
      \put(3533,4709){\makebox(0,0){\strut{}effective parameter 1D vs 2D}}%
    }%
    \gplgaddtomacro\gplfronttext{%
      \csname LTb\endcsname%
      \put(4939,1542){\makebox(0,0)[r]{\strut{}1D-t}}%
      \csname LTb\endcsname%
      \put(4939,1322){\makebox(0,0)[r]{\strut{}1D-V}}%
      \csname LTb\endcsname%
      \put(4939,1102){\makebox(0,0)[r]{\strut{}2D-t}}%
      \csname LTb\endcsname%
      \put(4939,882){\makebox(0,0)[r]{\strut{}2D-V}}%
    }%
    \gplbacktext
    \put(0,0){\includegraphics{effective-parameter-1Dvs2D}}%
    \gplfronttext
  \end{picture}%
\endgroup

%% file: img/trioncorrelation-U08-4x4-06-222-contour.tex
\begingroup
  \makeatletter
  \providecommand\color[2][]{%
    \GenericError{(gnuplot) \space\space\space\@spaces}{%
      Package color not loaded in conjunction with
      terminal option `colourtext'%
    }{See the gnuplot documentation for explanation.%
    }{Either use 'blacktext' in gnuplot or load the package
      color.sty in LaTeX.}%
    \renewcommand\color[2][]{}%
  }%
  \providecommand\includegraphics[2][]{%
    \GenericError{(gnuplot) \space\space\space\@spaces}{%
      Package graphicx or graphics not loaded%
    }{See the gnuplot documentation for explanation.%
    }{The gnuplot epslatex terminal needs graphicx.sty or graphics.sty.}%
    \renewcommand\includegraphics[2][]{}%
  }%
  \providecommand\rotatebox[2]{#2}%
  \@ifundefined{ifGPcolor}{%
    \newif\ifGPcolor
    \GPcolortrue
  }{}%
  \@ifundefined{ifGPblacktext}{%
    \newif\ifGPblacktext
    \GPblacktexttrue
  }{}%
  \let\gplgaddtomacro\g@addto@macro
  \gdef\gplbacktext{}%
  \gdef\gplfronttext{}%
  \makeatother
  \ifGPblacktext
    \def\colorrgb#1{}%
    \def\colorgray#1{}%
  \else
    \ifGPcolor
      \def\colorrgb#1{\color[rgb]{#1}}%
      \def\colorgray#1{\color[gray]{#1}}%
      \expandafter\def\csname LTw\endcsname{\color{white}}%
      \expandafter\def\csname LTb\endcsname{\color{black}}%
      \expandafter\def\csname LTa\endcsname{\color{black}}%
      \expandafter\def\csname LT0\endcsname{\color[rgb]{1,0,0}}%
      \expandafter\def\csname LT1\endcsname{\color[rgb]{0,1,0}}%
      \expandafter\def\csname LT2\endcsname{\color[rgb]{0,0,1}}%
      \expandafter\def\csname LT3\endcsname{\color[rgb]{1,0,1}}%
      \expandafter\def\csname LT4\endcsname{\color[rgb]{0,1,1}}%
      \expandafter\def\csname LT5\endcsname{\color[rgb]{1,1,0}}%
      \expandafter\def\csname LT6\endcsname{\color[rgb]{0,0,0}}%
      \expandafter\def\csname LT7\endcsname{\color[rgb]{1,0.3,0}}%
      \expandafter\def\csname LT8\endcsname{\color[rgb]{0.5,0.5,0.5}}%
    \else
      \def\colorrgb#1{\color{black}}%
      \def\colorgray#1{\color[gray]{#1}}%
      \expandafter\def\csname LTw\endcsname{\color{white}}%
      \expandafter\def\csname LTb\endcsname{\color{black}}%
      \expandafter\def\csname LTa\endcsname{\color{black}}%
      \expandafter\def\csname LT0\endcsname{\color{black}}%
      \expandafter\def\csname LT1\endcsname{\color{black}}%
      \expandafter\def\csname LT2\endcsname{\color{black}}%
      \expandafter\def\csname LT3\endcsname{\color{black}}%
      \expandafter\def\csname LT4\endcsname{\color{black}}%
      \expandafter\def\csname LT5\endcsname{\color{black}}%
      \expandafter\def\csname LT6\endcsname{\color{black}}%
      \expandafter\def\csname LT7\endcsname{\color{black}}%
      \expandafter\def\csname LT8\endcsname{\color{black}}%
    \fi
  \fi
  \setlength{\unitlength}{0.0500bp}%
  \begin{picture}(7200.00,5040.00)%
    \gplgaddtomacro\gplbacktext{%
      \csname LTb\endcsname%
      \put(1030,3288){\makebox(0,0)[r]{\strut{} 0}}%
      \put(1030,3545){\makebox(0,0)[r]{\strut{} 0.25}}%
      \put(1030,3802){\makebox(0,0)[r]{\strut{} 0.5}}%
      \put(1030,4059){\makebox(0,0)[r]{\strut{} 0.75}}%
      \put(1030,4316){\makebox(0,0)[r]{\strut{} 1}}%
      \put(1225,4599){\makebox(0,0){\strut{} 0}}%
      \put(2246,4599){\makebox(0,0){\strut{} 1}}%
      \put(3267,4599){\makebox(0,0){\strut{} 2}}%
      \put(4287,4599){\makebox(0,0){\strut{} 3}}%
      \put(5308,4599){\makebox(0,0){\strut{} 4}}%
      \put(6329,4599){\makebox(0,0){\strut{} 5}}%
      \put(128,3802){\rotatebox{-270}{\makebox(0,0){\strut{}$ n^{t}(x) n^{t}(0) $}}}%
      \put(3777,4928){\makebox(0,0){\strut{}$ d_{x} $}}%
    }%
    \gplgaddtomacro\gplfronttext{%
    }%
    \gplgaddtomacro\gplbacktext{%
    }%
    \gplgaddtomacro\gplfronttext{%
      \csname LTb\endcsname%
      \put(1799,103){\makebox(0,0){\strut{}0}}%
      \put(3419,103){\makebox(0,0){\strut{}1}}%
      \put(5039,103){\makebox(0,0){\strut{}2}}%
      \put(3419,-227){\makebox(0,0){\strut{}$d_{x}$}}%
      \put(735,879){\makebox(0,0)[r]{\strut{}0}}%
      \put(735,1696){\makebox(0,0)[r]{\strut{}1}}%
      \put(735,2513){\makebox(0,0)[r]{\strut{}2}}%
      \put(406,1696){\rotatebox{-270}{\makebox(0,0){\strut{}$d_{y}$}}}%
      \put(6535,470){\makebox(0,0)[l]{\strut{} 0}}%
      \put(6535,715){\makebox(0,0)[l]{\strut{} 0.02}}%
      \put(6535,960){\makebox(0,0)[l]{\strut{} 0.04}}%
      \put(6535,1205){\makebox(0,0)[l]{\strut{} 0.06}}%
      \put(6535,1450){\makebox(0,0)[l]{\strut{} 0.08}}%
      \put(6535,1695){\makebox(0,0)[l]{\strut{} 0.1}}%
      \put(6535,1940){\makebox(0,0)[l]{\strut{} 0.12}}%
      \put(6535,2185){\makebox(0,0)[l]{\strut{} 0.14}}%
      \put(6535,2430){\makebox(0,0)[l]{\strut{} 0.16}}%
      \put(6535,2675){\makebox(0,0)[l]{\strut{} 0.18}}%
      \put(6535,2920){\makebox(0,0)[l]{\strut{} 0.2}}%
    }%
    \gplbacktext
    \put(0,0){\includegraphics{trioncorrelation-U08-4x4-06-222-contour}}%
    \gplfronttext
  \end{picture}%
\endgroup

%% file: img/TrionCorrelationDiffDependence.tex
\begingroup
  \makeatletter
  \providecommand\color[2][]{%
    \GenericError{(gnuplot) \space\space\space\@spaces}{%
      Package color not loaded in conjunction with
      terminal option `colourtext'%
    }{See the gnuplot documentation for explanation.%
    }{Either use 'blacktext' in gnuplot or load the package
      color.sty in LaTeX.}%
    \renewcommand\color[2][]{}%
  }%
  \providecommand\includegraphics[2][]{%
    \GenericError{(gnuplot) \space\space\space\@spaces}{%
      Package graphicx or graphics not loaded%
    }{See the gnuplot documentation for explanation.%
    }{The gnuplot epslatex terminal needs graphicx.sty or graphics.sty.}%
    \renewcommand\includegraphics[2][]{}%
  }%
  \providecommand\rotatebox[2]{#2}%
  \@ifundefined{ifGPcolor}{%
    \newif\ifGPcolor
    \GPcolortrue
  }{}%
  \@ifundefined{ifGPblacktext}{%
    \newif\ifGPblacktext
    \GPblacktexttrue
  }{}%
  \let\gplgaddtomacro\g@addto@macro
  \gdef\gplbacktext{}%
  \gdef\gplfronttext{}%
  \makeatother
  \ifGPblacktext
    \def\colorrgb#1{}%
    \def\colorgray#1{}%
  \else
    \ifGPcolor
      \def\colorrgb#1{\color[rgb]{#1}}%
      \def\colorgray#1{\color[gray]{#1}}%
      \expandafter\def\csname LTw\endcsname{\color{white}}%
      \expandafter\def\csname LTb\endcsname{\color{black}}%
      \expandafter\def\csname LTa\endcsname{\color{black}}%
      \expandafter\def\csname LT0\endcsname{\color[rgb]{1,0,0}}%
      \expandafter\def\csname LT1\endcsname{\color[rgb]{0,1,0}}%
      \expandafter\def\csname LT2\endcsname{\color[rgb]{0,0,1}}%
      \expandafter\def\csname LT3\endcsname{\color[rgb]{1,0,1}}%
      \expandafter\def\csname LT4\endcsname{\color[rgb]{0,1,1}}%
      \expandafter\def\csname LT5\endcsname{\color[rgb]{1,1,0}}%
      \expandafter\def\csname LT6\endcsname{\color[rgb]{0,0,0}}%
      \expandafter\def\csname LT7\endcsname{\color[rgb]{1,0.3,0}}%
      \expandafter\def\csname LT8\endcsname{\color[rgb]{0.5,0.5,0.5}}%
    \else
      \def\colorrgb#1{\color{black}}%
      \def\colorgray#1{\color[gray]{#1}}%
      \expandafter\def\csname LTw\endcsname{\color{white}}%
      \expandafter\def\csname LTb\endcsname{\color{black}}%
      \expandafter\def\csname LTa\endcsname{\color{black}}%
      \expandafter\def\csname LT0\endcsname{\color{black}}%
      \expandafter\def\csname LT1\endcsname{\color{black}}%
      \expandafter\def\csname LT2\endcsname{\color{black}}%
      \expandafter\def\csname LT3\endcsname{\color{black}}%
      \expandafter\def\csname LT4\endcsname{\color{black}}%
      \expandafter\def\csname LT5\endcsname{\color{black}}%
      \expandafter\def\csname LT6\endcsname{\color{black}}%
      \expandafter\def\csname LT7\endcsname{\color{black}}%
      \expandafter\def\csname LT8\endcsname{\color{black}}%
    \fi
  \fi
  \setlength{\unitlength}{0.0500bp}%
  \begin{picture}(7200.00,5040.00)%
    \gplgaddtomacro\gplbacktext{%
      \csname LTb\endcsname%
      \put(1122,767){\makebox(0,0)[r]{\strut{} 0}}%
      \put(1122,1435){\makebox(0,0)[r]{\strut{} 0.005}}%
      \put(1122,2103){\makebox(0,0)[r]{\strut{} 0.01}}%
      \put(1122,2771){\makebox(0,0)[r]{\strut{} 0.015}}%
      \put(1122,3439){\makebox(0,0)[r]{\strut{} 0.02}}%
      \put(1122,4107){\makebox(0,0)[r]{\strut{} 0.025}}%
      \put(1122,4775){\makebox(0,0)[r]{\strut{} 0.03}}%
      \put(1317,484){\makebox(0,0){\strut{} 0}}%
      \put(2011,484){\makebox(0,0){\strut{} 10}}%
      \put(2705,484){\makebox(0,0){\strut{} 20}}%
      \put(3399,484){\makebox(0,0){\strut{} 30}}%
      \put(4093,484){\makebox(0,0){\strut{} 40}}%
      \put(4787,484){\makebox(0,0){\strut{} 50}}%
      \put(5481,484){\makebox(0,0){\strut{} 60}}%
      \put(6175,484){\makebox(0,0){\strut{} 70}}%
      \put(6869,484){\makebox(0,0){\strut{} 80}}%
      \put(4093,154){\makebox(0,0){\strut{}$U/t$}}%
    }%
    \gplgaddtomacro\gplfronttext{%
      \csname LTb\endcsname%
      \put(6014,4665){\makebox(0,0)[r]{\strut{}$ n^t(2,2) - n^f(2,2) $}}%
      \csname LTb\endcsname%
      \put(6014,4445){\makebox(0,0)[r]{\strut{}best fit}}%
      \csname LTb\endcsname%
      \put(6014,4225){\makebox(0,0)[r]{\strut{}1/U fit}}%
    }%
    \gplbacktext
    \put(0,0){\includegraphics{TrionCorrelationDiffDependence}}%
    \gplfronttext
  \end{picture}%
\endgroup

%% file: trionpaper.bbl
\begin{thebibliography}{99}
        \begin{thebibliography}{21}
\expandafter\ifx\csname natexlab\endcsname\relax\def\natexlab#1{#1}\fi
\expandafter\ifx\csname bibnamefont\endcsname\relax
  \def\bibnamefont#1{#1}\fi
\expandafter\ifx\csname bibfnamefont\endcsname\relax
  \def\bibfnamefont#1{#1}\fi
\expandafter\ifx\csname citenamefont\endcsname\relax
  \def\citenamefont#1{#1}\fi
\expandafter\ifx\csname url\endcsname\relax
  \def\url#1{\texttt{#1}}\fi
\expandafter\ifx\csname urlprefix\endcsname\relax\def\urlprefix{URL }\fi
\providecommand{\bibinfo}[2]{#2}
\providecommand{\eprint}[2][]{\url{#2}}

\bibitem[{\citenamefont{Bloch}(2008)}]{ImmanuelBloch02292008}
\bibinfo{author}{\bibfnamefont{I.}~\bibnamefont{Bloch}},
  \bibinfo{journal}{Science} \textbf{\bibinfo{volume}{319}},
  \bibinfo{pages}{1202} (\bibinfo{year}{2008}),
  \eprint{http://www.sciencemag.org/cgi/reprint/319/5867/1202.pdf},
  \urlprefix\url{http://www.sciencemag.org/cgi/content/abstract/319/5867/1202}.

\bibitem[{\citenamefont{Honerkamp and
  Hofstetter}(2004{\natexlab{a}})}]{PhysRevLett.92.170403}
\bibinfo{author}{\bibfnamefont{C.}~\bibnamefont{Honerkamp}} \bibnamefont{and}
  \bibinfo{author}{\bibfnamefont{W.}~\bibnamefont{Hofstetter}},
  \bibinfo{journal}{Phys. Rev. Lett.} \textbf{\bibinfo{volume}{92}},
  \bibinfo{pages}{170403} (\bibinfo{year}{2004}{\natexlab{a}}).

\bibitem[{\citenamefont{Honerkamp and
  Hofstetter}(2004{\natexlab{b}})}]{PhysRevB.70.094521}
\bibinfo{author}{\bibfnamefont{C.}~\bibnamefont{Honerkamp}} \bibnamefont{and}
  \bibinfo{author}{\bibfnamefont{W.}~\bibnamefont{Hofstetter}},
  \bibinfo{journal}{Phys. Rev. B} \textbf{\bibinfo{volume}{70}},
  \bibinfo{pages}{094521} (\bibinfo{year}{2004}{\natexlab{b}}).

\bibitem[{\citenamefont{Rapp et~al.}(2007)\citenamefont{Rapp, Zar\'and,
  Honerkamp, and Hofstetter}}]{PhysRevLett.98.160405}
\bibinfo{author}{\bibfnamefont{A.}~\bibnamefont{Rapp}},
  \bibinfo{author}{\bibfnamefont{G.}~\bibnamefont{Zar\'and}},
  \bibinfo{author}{\bibfnamefont{C.}~\bibnamefont{Honerkamp}},
  \bibnamefont{and}
  \bibinfo{author}{\bibfnamefont{W.}~\bibnamefont{Hofstetter}},
  \bibinfo{journal}{Phys. Rev. Lett.} \textbf{\bibinfo{volume}{98}},
  \bibinfo{pages}{160405} (\bibinfo{year}{2007}).

\bibitem[{\citenamefont{Rapp et~al.}(2008)\citenamefont{Rapp, Hofstetter, and
  Zar\'and}}]{PhysRevB.77.144520}
\bibinfo{author}{\bibfnamefont{A.}~\bibnamefont{Rapp}},
  \bibinfo{author}{\bibfnamefont{W.}~\bibnamefont{Hofstetter}},
  \bibnamefont{and} \bibinfo{author}{\bibfnamefont{G.}~\bibnamefont{Zar\'and}},
  \bibinfo{journal}{Phys. Rev. B} \textbf{\bibinfo{volume}{77}},
  \bibinfo{pages}{144520} (\bibinfo{year}{2008}).

\bibitem[{\citenamefont{Cherng et~al.}(2007)\citenamefont{Cherng, Refael, and
  Demler}}]{PhysRevLett.99.130406}
\bibinfo{author}{\bibfnamefont{R.~W.} \bibnamefont{Cherng}},
  \bibinfo{author}{\bibfnamefont{G.}~\bibnamefont{Refael}}, \bibnamefont{and}
  \bibinfo{author}{\bibfnamefont{E.}~\bibnamefont{Demler}},
  \bibinfo{journal}{Phys. Rev. Lett.} \textbf{\bibinfo{volume}{99}},
  \bibinfo{pages}{130406} (\bibinfo{year}{2007}).

\bibitem[{\citenamefont{Capponi et~al.}(2008)\citenamefont{Capponi, Roux,
  Lecheminant, Azaria, Boulat, and White}}]{PhysRevA.77.013624}
\bibinfo{author}{\bibfnamefont{S.}~\bibnamefont{Capponi}},
  \bibinfo{author}{\bibfnamefont{G.}~\bibnamefont{Roux}},
  \bibinfo{author}{\bibfnamefont{P.}~\bibnamefont{Lecheminant}},
  \bibinfo{author}{\bibfnamefont{P.}~\bibnamefont{Azaria}},
  \bibinfo{author}{\bibfnamefont{E.}~\bibnamefont{Boulat}}, \bibnamefont{and}
  \bibinfo{author}{\bibfnamefont{S.~R.} \bibnamefont{White}},
  \bibinfo{journal}{Phys. Rev. A} \textbf{\bibinfo{volume}{77}},
  \bibinfo{pages}{013624} (\bibinfo{year}{2008}).

\bibitem[{\citenamefont{Hertzberg et~al.}(2008)\citenamefont{Hertzberg,
  Tegmark, and Wilczek}}]{Hertzberg:2008wr}
\bibinfo{author}{\bibfnamefont{M.~P.} \bibnamefont{Hertzberg}},
  \bibinfo{author}{\bibfnamefont{M.}~\bibnamefont{Tegmark}}, \bibnamefont{and}
  \bibinfo{author}{\bibfnamefont{F.}~\bibnamefont{Wilczek}},
  \bibinfo{journal}{Phys. Rev.} \textbf{\bibinfo{volume}{D78}},
  \bibinfo{pages}{083507} (\bibinfo{year}{2008}), \eprint{0807.1726}.

\bibitem[{\citenamefont{Inaba and Suga}(2009)}]{PhysRevA.80.041602}
\bibinfo{author}{\bibfnamefont{K.}~\bibnamefont{Inaba}} \bibnamefont{and}
  \bibinfo{author}{\bibfnamefont{S.~I.} \bibnamefont{Suga}},
  \bibinfo{journal}{Phys. Rev. A} \textbf{\bibinfo{volume}{80}},
  \bibinfo{pages}{041602(R)} (\bibinfo{year}{2009}).

\bibitem[{\citenamefont{Molina et~al.}(2009)\citenamefont{Molina, Dukelsky, and
  Schmitteckert}}]{PhysRevA.80.013616}
\bibinfo{author}{\bibfnamefont{R.~A.} \bibnamefont{Molina}},
  \bibinfo{author}{\bibfnamefont{J.}~\bibnamefont{Dukelsky}}, \bibnamefont{and}
  \bibinfo{author}{\bibfnamefont{P.}~\bibnamefont{Schmitteckert}},
  \bibinfo{journal}{Phys. Rev. A} \textbf{\bibinfo{volume}{80}},
  \bibinfo{pages}{013616} (\bibinfo{year}{2009}).

\bibitem[{\citenamefont{Luscher and Laeuchli}(2009)}]{luscher-2009}
\bibinfo{author}{\bibfnamefont{A.}~\bibnamefont{Luscher}} \bibnamefont{and}
  \bibinfo{author}{\bibfnamefont{A.}~\bibnamefont{Laeuchli}},
  \emph{\bibinfo{title}{Imbalanced thee-component fermi gas with attractive
  interactions: Multiple fflo-pairing, bose-fermi and fermi-fermi mixtures
  versus collapse and phase separation}} (\bibinfo{year}{2009}),
  \urlprefix\url{http://www.citebase.org/abstract?id=oai:arXiv.org:0906.0768}.

\bibitem[{\citenamefont{Errea et~al.}(2009)\citenamefont{Errea, Dukelsky, and
  Ortiz}}]{PhysRevA.79.051603}
\bibinfo{author}{\bibfnamefont{B.}~\bibnamefont{Errea}},
  \bibinfo{author}{\bibfnamefont{J.}~\bibnamefont{Dukelsky}}, \bibnamefont{and}
  \bibinfo{author}{\bibfnamefont{G.}~\bibnamefont{Ortiz}},
  \bibinfo{journal}{Phys. Rev. A} \textbf{\bibinfo{volume}{79}},
  \bibinfo{pages}{051603(R)} (\bibinfo{year}{2009}).

\bibitem[{\citenamefont{Ottenstein et~al.}(2008)\citenamefont{Ottenstein,
  Lompe, Kohnen, Wenz, and Jochim}}]{PhysRevLett.101.203202}
\bibinfo{author}{\bibfnamefont{T.~B.} \bibnamefont{Ottenstein}},
  \bibinfo{author}{\bibfnamefont{T.}~\bibnamefont{Lompe}},
  \bibinfo{author}{\bibfnamefont{M.}~\bibnamefont{Kohnen}},
  \bibinfo{author}{\bibfnamefont{A.~N.} \bibnamefont{Wenz}}, \bibnamefont{and}
  \bibinfo{author}{\bibfnamefont{S.}~\bibnamefont{Jochim}},
  \bibinfo{journal}{Phys. Rev. Lett.} \textbf{\bibinfo{volume}{101}},
  \bibinfo{pages}{203202} (\bibinfo{year}{2008}).

\bibitem[{\citenamefont{Wille et~al.}(2008)\citenamefont{Wille, Spiegelhalder,
  Kerner, Naik, Trenkwalder, Hendl, Schreck, Grimm, Tiecke, Walraven
  et~al.}}]{PhysRevLett.100.053201}
\bibinfo{author}{\bibfnamefont{E.}~\bibnamefont{Wille}},
  \bibinfo{author}{\bibfnamefont{F.~M.} \bibnamefont{Spiegelhalder}},
  \bibinfo{author}{\bibfnamefont{G.}~\bibnamefont{Kerner}},
  \bibinfo{author}{\bibfnamefont{D.}~\bibnamefont{Naik}},
  \bibinfo{author}{\bibfnamefont{A.}~\bibnamefont{Trenkwalder}},
  \bibinfo{author}{\bibfnamefont{G.}~\bibnamefont{Hendl}},
  \bibinfo{author}{\bibfnamefont{F.}~\bibnamefont{Schreck}},
  \bibinfo{author}{\bibfnamefont{R.}~\bibnamefont{Grimm}},
  \bibinfo{author}{\bibfnamefont{T.~G.} \bibnamefont{Tiecke}},
  \bibinfo{author}{\bibfnamefont{J.~T.~M.} \bibnamefont{Walraven}},
  \bibnamefont{et~al.}, \bibinfo{journal}{Phys. Rev. Lett.}
  \textbf{\bibinfo{volume}{100}}, \bibinfo{pages}{053201}
  (\bibinfo{year}{2008}).

\bibitem[{\citenamefont{Fukuhara et~al.}(2007)\citenamefont{Fukuhara, Takasu,
  Kumakura, and Takahashi}}]{PhysRevLett.98.030401}
\bibinfo{author}{\bibfnamefont{T.}~\bibnamefont{Fukuhara}},
  \bibinfo{author}{\bibfnamefont{Y.}~\bibnamefont{Takasu}},
  \bibinfo{author}{\bibfnamefont{M.}~\bibnamefont{Kumakura}}, \bibnamefont{and}
  \bibinfo{author}{\bibfnamefont{Y.}~\bibnamefont{Takahashi}},
  \bibinfo{journal}{Phys. Rev. Lett.} \textbf{\bibinfo{volume}{98}},
  \bibinfo{pages}{030401} (\bibinfo{year}{2007}).

\bibitem[{\citenamefont{Blume et~al.}(2008)\citenamefont{Blume, Rittenhouse,
  von Stecher, and Greene}}]{PhysRevA.77.033627}
\bibinfo{author}{\bibfnamefont{D.}~\bibnamefont{Blume}},
  \bibinfo{author}{\bibfnamefont{S.~T.} \bibnamefont{Rittenhouse}},
  \bibinfo{author}{\bibfnamefont{J.}~\bibnamefont{von Stecher}},
  \bibnamefont{and} \bibinfo{author}{\bibfnamefont{C.~H.}
  \bibnamefont{Greene}}, \bibinfo{journal}{Phys. Rev. A}
  \textbf{\bibinfo{volume}{77}}, \bibinfo{pages}{033627}
  (\bibinfo{year}{2008}).

\bibitem[{\citenamefont{Floerchinger et~al.}(2009)\citenamefont{Floerchinger,
  Schmidt, and Wetterich}}]{PhysRevA.79.053633}
\bibinfo{author}{\bibfnamefont{S.}~\bibnamefont{Floerchinger}},
  \bibinfo{author}{\bibfnamefont{R.}~\bibnamefont{Schmidt}}, \bibnamefont{and}
  \bibinfo{author}{\bibfnamefont{C.}~\bibnamefont{Wetterich}},
  \bibinfo{journal}{Phys. Rev. A} \textbf{\bibinfo{volume}{79}},
  \bibinfo{pages}{053633} (\bibinfo{year}{2009}).

\bibitem[{\citenamefont{Huckans et~al.}(2009)\citenamefont{Huckans, Williams,
  Hazlett, Stites, and O'Hara}}]{PhysRevLett.102.165302}
\bibinfo{author}{\bibfnamefont{J.~H.} \bibnamefont{Huckans}},
  \bibinfo{author}{\bibfnamefont{J.~R.} \bibnamefont{Williams}},
  \bibinfo{author}{\bibfnamefont{E.~L.} \bibnamefont{Hazlett}},
  \bibinfo{author}{\bibfnamefont{R.~W.} \bibnamefont{Stites}},
  \bibnamefont{and} \bibinfo{author}{\bibfnamefont{K.~M.}
  \bibnamefont{O'Hara}}, \bibinfo{journal}{Phys. Rev. Lett.}
  \textbf{\bibinfo{volume}{102}}, \bibinfo{pages}{165302}
  (\bibinfo{year}{2009}).

\bibitem[{\citenamefont{Paananen et~al.}(2006)\citenamefont{Paananen,
  Martikainen, and T\"orm\"a}}]{PhysRevA.73.053606}
\bibinfo{author}{\bibfnamefont{T.}~\bibnamefont{Paananen}},
  \bibinfo{author}{\bibfnamefont{J.-P.} \bibnamefont{Martikainen}},
  \bibnamefont{and}
  \bibinfo{author}{\bibfnamefont{P.}~\bibnamefont{T\"orm\"a}},
  \bibinfo{journal}{Phys. Rev. A} \textbf{\bibinfo{volume}{73}},
  \bibinfo{pages}{053606} (\bibinfo{year}{2006}).

\bibitem[{\citenamefont{Toke and Hofstetter}()}]{toke}
\bibinfo{author}{\bibfnamefont{C.}~\bibnamefont{Toke}} \bibnamefont{and}
  \bibinfo{author}{\bibfnamefont{W.}~\bibnamefont{Hofstetter}},
  \urlprefix\url{unpublished}.

\bibitem[{\citenamefont{Azaria et~al.}(2009)\citenamefont{Azaria, Capponi, and
  Lecheminant}}]{PhysRevA.80.041604}
\bibinfo{author}{\bibfnamefont{P.}~\bibnamefont{Azaria}},
  \bibinfo{author}{\bibfnamefont{S.}~\bibnamefont{Capponi}}, \bibnamefont{and}
  \bibinfo{author}{\bibfnamefont{P.}~\bibnamefont{Lecheminant}},
  \bibinfo{journal}{Phys. Rev. A} \textbf{\bibinfo{volume}{80}},
  \bibinfo{pages}{041604(R)} (\bibinfo{year}{2009}).

\end{thebibliography}
